\documentclass[12pt]{article}
\title{On a three-dimensional symmetric Ising tetrahedron, and contributions
to the theory of the dilogarithm and Clausen functions}
\author{Mark W. Coffey\\
Department of Physics\\
Colorado School of Mines\\
Golden, CO  80401\\
(Received $\mbox{~~~~~~~~~~~~~~~~~~~~~~~~~~~~~~~2008}$)}
\date{January 18, 2008}
\begin{document}
\maketitle
\begin{center}

\baselineskip=25 pt
\begin{abstract}

Perturbative quantum field theory for the Ising model at the three-loop level
yields a tetrahedral Feynman diagram $C(a,b)$ with masses $a$ and $b$ and
four other lines with unit mass.  The completely symmetric tetrahedron
$C^{\mbox{Tet}} \equiv C(1,1)$ has been of interest from many points of view,
with several representations and conjectures having been given in the literature.
We prove a conjectured exponentially fast convergent sum for $C(1,1)$, as well
as a previously empirical relation for $C(1,1)$ 
as a remarkable difference of Clausen function values.  Our presentation 
includes Propositions extending the theory of the dilogarithm Li$_2$ and Clausen 
Cl$_2$ functions, as well as their relation to other special functions of
mathematical physics.
The results strengthen connections between Feynman diagram integrals, 
volumes in hyperbolic space, number theory, and special functions and numbers, specifically including dilogarithms, Clausen function values, and harmonic
numbers.

\end{abstract}

\pagebreak
\baselineskip=15pt
\centerline{\bf Key words and phrases}
\medskip

\noindent
Feynman diagram, Ising tetrahedron,
Clausen function, dilogarithm function, Legendre chi function, Lerch zeta 
function, hypergeometric function, digamma function, polygamma function, 
harmonic numbers, generalized harmonic numbers, Lobachevskiy function 
\end{center}

\bigskip
\centerline{\bf AMS classification numbers}
33B30, 33B15, 33C20 

\bigskip
\centerline{\bf PACS classification numbers}
02.30.Gp, 02.30.-f, 02.10.De


\baselineskip=25pt
\pagebreak
\medskip
\centerline{\bf Introduction}
\medskip

In Refs. \cite{broadhurst,broad2}, Broadhurst considered a dispersive integral coming
from a tetrahedral Feynman diagram of $\phi^4$ field theory at the three-loop
level.  The integral $C(a,b)$ corresponds to the tetrahedron with non-adjacent
lines with masses $a$ and $b$ and the other four lines with unit mass.  The
symmetry $C(a,b)=C(b,a)$ holds, as well as the limit $\lim_{a \to \infty} a
C(a,b)=0$, the latter property due to a superconvergence relation.  The 
totally symmetric tetrahedron, $C^{\mbox{Tet}} \equiv C(1,1)$, not only seems
to be expressible in terms of the dilogarithm function Li$_2$, but is 
conjectured on the basis of the lattice reduction program PSLQ to be given
by a remarkably compact difference of Clausen function values. 

The particular Clausen function Cl$_2$ can be defined by (e.g., \cite{lewin})
$$\mbox{Cl}_2(\theta)\equiv -\int_0^\theta\ln\left|2 \sin {t \over 2}\right|dt
=\int_0^1 \tan^{-1}\left({{x \sin \theta} \over {1-x\cos \theta}}\right)
{{dx} \over x}$$
$$=-\sin \theta \int_0^1 {{\ln x} \over {x^2-2x\cos \theta +1}}dx
=\sum_{n=1}^\infty {{\sin(n \theta)} \over n^2}.  \eqno(1)$$
When $\theta$ is a rational multiple of $\pi$ it is known that Cl$_2(\theta)$
may be written in terms of the trigamma and sine functions \cite{ded,grosjean}.
With the definition $\alpha \equiv \sin^{-1}(1/3)$, the conjectured value of
$C^{\mbox{Tet}}$ is given by
$$C(1,1)\stackrel{?}{=}2^{5/2}[\mbox{Cl}_2(4\alpha)-\mbox{Cl}_2(2\alpha)].  
\eqno(2)$$
While numerical calculation confirms Eq. (2) to high precision, this relation 
had remained unproven until now, and we use the $\stackrel{?}{=}$ notation
to make this readily apparent.

In addition, also with the aid of the PSLQ program, the expression for
$C^{\mbox{Tet}}$ was transformed into an exponentially convergent sum
\cite{broadhurst}.  This sum, specified in Eq. (3) below, is shown in our
first Proposition to be given exactly as the difference of the Cl$_2$ values
of Eq. (2).  
In our final Proposition 10 we demonstrate that (2) does hold as an equality.
Although we have also reduced the verification of (2) to the evaluation of a
sole type of integral at particular parametric values, as discussed in the
concluding section, we base our proof on previously obtained results in
atomic physics \cite{harris,remiddi}.

The conjectural relations of Broadhurst \cite{broadhurst} together with
our previous study \cite{coffey07} of a certain $\ln \tan$ integral of quantum
field theory, provide significant portions of our motivation for the present
investigation.  We are also interested to advance the state of the art 
in certain areas of integration and special function theory.  We focus
on certain harmonic and generalized harmonic number sums, the highly 
related Clausen Cl$_2$, dilogarithm Li$_2$, Log-sine integral Ls$_2$,
and Lobachevskiy $L$ functions, and their connections with other special 
functions of mathematical physics.  

Following Proposition 1 and its proof, and various mentioned alternative
methods of proof, we consider extended families of related harmonic number
sums.  We then give series and integral representations of Cl$_2$ in terms
of integrals of Legendre polynomials and otherwise.  We develop and apply
the exponential generating function $I(x,u)$ of Log-sine integrals Ls$_n$.
We then give important integrals expressible in terms of Clausen or
Lobachevskiy function values.  
Readers interested specifically in the massive three-loop Ising tetrahedron
$C^{\mbox{Tet}}$ may refer to Proposition 1, Appendices A and C, Proposition 7(b),
Corollary 9, and Proposition 10.  

\medskip
\centerline{\bf A difference of Clausen Cl$_2$ values}
\medskip

Let $H_n=\sum_{k=1}^n {1 \over k}$ be the $n$th harmonic number and define
$\omega \equiv \tan^{-1}(1/2\sqrt{2})$.  Then we have
{\newline \bf Proposition 1}.  We have
$$C(1,1) \equiv \sum_{n=0}^\infty {{(-1/2)^{3n}} \over {n+1/2}}\left[{1 \over
{n+1/2}}-3(\ln 2+H_n)\right]=4\sqrt{2}[\mbox{Cl}_2(4\omega)-\mbox{Cl}_2(2\omega)].
\eqno(3)$$

The proof uses two Lemmas that analytically evaluate component sums of Eq. (3).

Put $\theta_+ \equiv -\tan^{-1}(4\sqrt{2}/7)$ and
$$S(\alpha,\beta) \equiv \sum_{n=0}^\infty {{(-1/\beta)^{3n}} \over {n+1/\alpha}}
H_n,  \eqno(4)$$
with $|\beta|>1$.  Note that in this equation the sum may just as well start
at $n=1$, since $H_0=0$.  
{\newline \bf Lemma 1}.  We have
$$S(2,2)=4\sqrt{2}[\mbox{Cl}_2(\theta_+ +\pi)+ \theta_+ \ln 2]. \eqno(5)$$

Proof. We prove Lemma 1 and then describe six other methods of proof.
We commence with general integral representations of $S(\alpha,\beta)$ and
then specialize to $S(2,2)$.  We write $H_n=\psi(n+1)+\gamma$ where $\gamma$ is
the Euler constant and $\psi$ the digamma function, and then use an integral
representation of the latter function \cite{grad} (p. 943):
$$S(\alpha,\beta)=\sum_{n=0}^\infty {{(-1/\beta)^{3n}} \over {n+1/\alpha}}
[\psi(n+1)+\gamma]=\sum_{n=0}^\infty {{(-1/\beta)^{3n}} \over {n+1/\alpha}}
\int_0^1 {{t^n-1} \over {t-1}}dt.  \eqno(6)$$
The integral is absolutely convergent, we may interchange integration and
summation, and we first obtain
$$S(\alpha,\beta)=\alpha\int_0^1 \left[_2F_1\left(1,{1 \over \alpha};1+{1 \over
\alpha};-{t \over \beta^3}\right)-_2F_1\left(1,{1 \over \alpha};1+{1 \over
\alpha};-{1 \over \beta^3}\right)\right]{{dt} \over {t-1}}, \eqno(7)$$
where $_2F_1$ is the Gauss hypergeometric function.  Equation (7) immediately
follows from Eq. (6) because we have the ratio of Pochhammer symbols
$(1/\alpha)_j/(1+1/\alpha)_j=1/\alpha (j+1/\alpha)$.  Owing to the relation
\cite{grad} (p. 950)
$$B_x(p,q) \equiv \int_0^x t^{p-1}(1-t)^{q-1}dt={x^p \over p} ~_2F_1(p,1-q;p+1;
x), \eqno(8)$$
where $B_x(p,q)$ is the incomplete Beta function, we may alternatively write
$$S(\alpha,\beta)=(-\beta)^{1/\alpha}\int_0^1\left[t^{-1/\alpha}B_{-t/\beta^3}
\left({1 \over \alpha},0\right)-B_{-1/\beta^3}\left({1 \over \alpha},0 \right)
\right]  {{dt} \over {t-1}}.  \eqno(9)$$

We now specialize to $\alpha=\beta=2$ and use the well known relation
\cite{grad} (p. 1042) 
$$_2F_1\left({1 \over 2},1;{3 \over 2};-z^2\right)={1 \over z}\tan^{-1}z,
\eqno(10)$$
where $\tan^{-1}z = \cot^{-1}(1/z)$ for $z>0$, obtaining
$$S(2,2)=4\sqrt{2}\int_0^1\left[-\cot^{-1}(2\sqrt{2})+{1 \over \sqrt{t}}
\cot^{-1}\left({{2\sqrt{2}} \over \sqrt{t}}\right)\right] {{dt} \over {t-1}}$$
$$=64\sqrt{2}\int_0^{1/2\sqrt{2}}\left[-\cot^{-1}(2\sqrt{2})+{1 \over {2\sqrt{2}u}}
\cot^{-1}\left({1 \over u}\right)\right] {{udu} \over {(8u^2-1)}}.  \eqno(11)$$
By writing $\tan^{-1}u=(1/2i)\ln[(iu-1)/(iu+1)]$ and writing $1/(8u^2-1)$ in
terms of partial fractions, the integral can be evaluated in terms of the
dilogarithm function Li$_2$,
$$S(2,2)=2\sqrt{2}\left\{2\theta_+ \ln 2 -i[\mbox{Li}_2(-e^{i\theta_+})-
\mbox{Li}_2(-e^{-i\theta_+})]\right\}.  \eqno(12)$$
We recall that the difference of such dilogarithms may be written in terms
of the Clausen function Cl$_2$:
$$\mbox{Li}_2(e^{i\theta})-\mbox{Li}_2(e^{-i\theta}) = 2i \mbox{Cl}_2(\theta).
\eqno(13)$$
Lemma 1 follows.

Method 2.  The generating function of harmonic numbers,
$$(1-z)\sum_{n=1}^\infty H_n z^n =-\ln(1-z), \eqno(14)$$
may be used to write
$$\sum_{n=1}^\infty H_n \int_0^w z^{n+1/\alpha-1}dz=\sum_{n=1}^\infty {H_n \over {n+1/\alpha}} w^{n+1/\alpha}$$
$$=-\int_0^w z^{1/\alpha-1} {{\ln(1-z)} \over {1-z}}dz.  \eqno(15)$$
Performing the integral at $\alpha=2$, the result may be written in terms of
dilogarithms, 
$$\int_0^w {{\ln(1-z)} \over {\sqrt{z}(1-z)}}dz=\mbox{Li}_2\left[{1 \over 2}
(1-\sqrt{w})\right]-\mbox{Li}_2\left[{1 \over 2}(1+\sqrt{w})\right]+{1 \over 2}
\left[\ln^2(\sqrt{w}-1)-\ln^2(\sqrt{w}+1)\right]$$
$$+\ln 2(1-w) [\ln(\sqrt{w}+1)-\ln(\sqrt{w}-1)]+i\pi \ln(1+\sqrt{w})
+{\pi \over 2}(\pi + 2i\ln 2). \eqno(16)$$  
This result may be easily confirmed by differentiation with respect to $w$, given
Li$_2(z)=-\int_0^z \ln(1-t)(dt/t)$.  Again $S(2,2)$ can be determined.

Equivalently, we may write
$$\sum_{n=1}^\infty {H_n \over {n+1/\alpha}} w^n
={w^{-1/\alpha} \over 2}\int_0^w z^{1/\alpha-2} \ln^2(1-z)dz, \eqno(17)$$ 
that at $w=-1/\beta^3$ gives $S(\alpha,\beta)$.  This equation follows from
integration by parts on Eq. (15).  Alternatively, it follows from integrating
the generating function relation (14), so that
$$\sum_{n=1}^\infty H_nz^{n+1} = {1 \over 2} \ln^2(1-z).  \eqno(18)$$ 

Method 3.  We may begin with a method of Nielsen \cite{nielsen}, \cite{lewin}
(p. 22), wherein if a function $f$ has a power series development
$$f(x)=\sum_{j=0}^\infty a_j x^j, \eqno(19)$$
then
$$\sum_{n=1}^\infty {a_{n-1} \over n^2}H_n x^n = {\pi^2 \over 6}\int_0^x f(t) dt
-x\int_0^1 f(tx) \mbox{Li}_2(t)dt.  \eqno(20)$$
Here we put $a_{n-1}=n^2/(n+1/2)$, so that we obtain   
$$f(x)={1 \over {2x^{3/2}}}\left[{{\sqrt{x}(3x-1)} \over {(x-1)^2}} + \tanh^{-1}
\sqrt{x}\right],  \eqno(21)$$
and
$$\int_0^x f(t) dt={1 \over {1-x}}-{1 \over \sqrt{x}}\tanh^{-1} \sqrt{x}.  \eqno(22)$$

Method 4.  We may write $S(\alpha,\beta)$ as
$$S(\alpha,\beta)=\sum_{n=0}^\infty (-1/\beta)^{3n} H_n \int_0^\infty e^{-(n+1/\alpha)t} dt, \eqno(23)$$
again with an absolutely convergent integral.  Interchanging summation and
integration and making use of the generating function (14) we have
$$S(\alpha,\beta)=-\beta^3\int_0^\infty {e^{-t/\alpha} \over {e^{-t}+\beta^3}}
\ln[\beta^{-3}(e^{-t}+\beta^3)]dt$$
$$=-\beta^3\int_0^1 {u^{1/\alpha} \over {(u+\beta^3)}}\ln(1+\beta^{-3}u) {{du}
\over u}.  \eqno(24)$$
At $\alpha=\beta=2$ the integral may be performed and we obtain an equivalent
form
$$S(2,2)=\sqrt{2}\left\{-5(\pi-2\tan^{-1}(2\sqrt{2}))\ln 2+4\omega\ln(8/3)\right.$$
$$\left. + 2i\left[\mbox{Li}_2\left[{1 \over 8}\left(4-i\sqrt{2}\right)\right]-
\mbox{Li}_2\left[{1 \over 8}\left(4+i\sqrt{2}\right)\right]\right] \right\}.
\eqno(25)$$
It is also possible to expand the logarithm in Taylor series in Eq. (24), 
integrate, and thereby obtain summation representations in terms of sums over
hypergeometric functions.  However, we omit these developments.

Method 5 uses a summatory result of Ramanujan \cite{berndt}, and is given in
Appendix A.  Method 6 presented in Appendix C gives the general result for
$S(2,\beta)$ in terms of Cl$_2$.

Method 7 is to use an integral representation with free parameter $t$,
$$H_n=\ln t-(n+1)t^{n+1}\int_0^\infty {{\ln x ~dx} \over {(x+t)^{n+2}}}, ~~~~~~
n > -1, ~~~~\mbox{Re} ~t >0.  $$
This equation may be thought of as a $t \neq 1$ extension of Eq. (C.11).
Upon inserting this representation into the definition (4) we find
$$S(\alpha,\beta)=\alpha \ln t ~_2F_1\left(1,{1 \over \alpha};1+{1 \over \alpha};
-{1 \over \beta^3}\right)-\left\{ {\beta^3 \over 2}\left[\ln^2 {\beta^3 \over
{t(\beta^3+1)}}-\ln^2 {1 \over t}\right] \right.$$
$$\left. +(\alpha-1)t\int_0^\infty {{\ln x} \over {(x+t)^2}}
~_2F_1\left(1,{1 \over \alpha};1+{1 \over \alpha};-{t \over {\beta^3(x+t)}}\right)
dx \right\}.$$
When $\alpha=2$, Eq. (10) again applies, and at least for $\beta=2$ and $t=1$,
the integration in this equation may be performed in terms of dilogarithms,
leading again to Clausen function values.

For the next Lemma, we let $\Phi(z,s,a)$ be the Lerch zeta function (e.g., 
\cite{sri}, p. 121), and
$$\chi_s(z)=\sum_{n=0}^\infty {z^{2n+1} \over {(2n+1)^s}}, ~~~~~~|z| \leq 1,
~~~~~~\mbox{Re} ~s>1, \eqno(26)$$
be the Legendre chi function.  We have
{\newline \bf Lemma 2}.  
$$\sum_{n=0}^\infty {{(-1/2)^{3n}} \over {(n+1/2)^2}}=\Phi\left(-{1 \over 8},
2,{1 \over 2}\right)=4\sqrt{2}\left[2\omega \ln\left({1 \over {2\sqrt{2}}}
\right)+\mbox{Cl}_2(2\omega)-\mbox{Cl}_2(2\omega+\pi)\right].  \eqno(27)$$

Proof.  We have
$$\sum_{n=0}^\infty {{(-1/2)^{3n}} \over {(n+1/2)^2}}=\Phi\left(-{1 \over 8},
2,{1 \over 2}\right)={4 \over z}\chi_2(z)$$
$$={2 \over z}[\mbox{Li}_2(z)-\mbox{Li}_2(-z)], \eqno(28)$$
where $z=i/2\sqrt{2}$ and we used \cite{sri} (p. 108).  Now the difference of
dilogarithms in this equation is pure imaginary, as it must be, and it suffices
to take the imaginary part of the particular values Li$_2(\pm z)$, using 
the relation 
$$\mbox{Im Li}_2(re^{i\theta})=\omega \ln r+{1 \over 2}\left[\mbox{Cl}_2(2\omega)
-\mbox{Cl}_2(2\omega+2\theta) + \mbox{Cl}_2(2\theta)\right],  \eqno(29)$$
where
$$\omega=\tan^{-1}\left({{r \sin \theta} \over {1-r\cos \theta}}\right).
\eqno(30)$$
We apply Eqs. (29) and (30) at $\theta=\pm \pi/2$, note that Cl$_2(\pi)=0$
and Cl$_2(\pm \theta)=\pm \mbox{Cl}_2(\theta)$, and Lemma 2 follows.

{\bf Remark}.  Relevant to Lemma 2 and to further developments is the integral
representation of the $\chi_2$ function \cite{hansen} (p. 54)
$$\int_0^x {1 \over t}\tanh^{-1} t ~dt={1 \over 2}[\mbox{Li}_2(x)-\mbox{Li}_2(-x)]
=\sum_{k=0}^\infty {x^{2k+1} \over {(2k+1)^2}}, ~~~~~~|x| \leq 1.  \eqno(31)$$
By a change of variable, we have
$$\chi_2(x)= {1 \over 2}\int_0^{x^2} \tanh^{-1} \sqrt{u} ~{{du} \over u}.  \eqno(32)$$

We quickly present another way to prove Lemma 2.  By observing that
$$\int_0^y x^{p+k-1} \ln x ~dx=y^{k+p} {{[-1+(k+p)\ln y]} \over {(k+p)^2}}, ~~~~~~
\mbox{Re} ~(p+k)>0, \eqno(33)$$
we have at $y=-1/8$, $p=1/2$, and summing over $k$ from $0$ to $\infty$,
$$\int_0^{-1/8} {1 \over {(1-x)}}{{\ln x} \over \sqrt{x}}~dx=\left(-{1 \over 8}\right)^{1/2} \sum_{k=0}^\infty \left(-{1 \over 8}\right)^k {{[-1+(k+1/2)\ln(-1/8)]}
\over {(k+1/2)^2}}$$
$$~~~~~~~~~~~~~~~~~~~~~~ = {i \over {2 \sqrt{2}}}\left[\Phi\left(-{1 \over 8},2, {1 \over 2}\right)+\ln\left(-{1 \over 8}\right)4\sqrt{2}\omega\right]. \eqno(34)$$
The integral on the left side of this equation may be evaluated in terms of
the difference of dilogarithms in Eq. (28) and hence Lemma 2 may be rederived.

In Appendix B, we gather other relations pertinent to the Lerch zeta function,
and to the value $\Phi(-1/8,2,1/2)$ in particular.

Proof of Proposition 1.  Similarly to a step in the proof of Lemma 1, we have
the simple subsum of Eq. (1)
$$\sum_{n=0}^\infty {{(-1/2)^{3n}} \over {n+1/2}}=2(2\sqrt{2} \omega-1).  \eqno(35)$$
We then combine the results of Eqs. (5), (27), and (35) to form $C(1,1)$.
As $\sin 2n \omega=-\sin n\theta_+$ for integers $n$, we have Cl$_2(2\omega+\pi)=-\mbox{Cl}_2 (\theta_+ +\pi)$.  
Upon simplification, we obtain
$$C(1,1)=4\sqrt{2}[\mbox{Cl}_2(2\omega)+2 \mbox{Cl}_2(2\omega+\pi)].  \eqno(36)$$
Finally, by use of the duplication formula satisfied by the Clausen function,
$${1 \over 2}\mbox{Cl}_2(2\theta)=\mbox{Cl}_2(\theta)-\mbox{Cl}_2(\pi-\theta),
\eqno(37)$$
we obtain the Proposition.

{\bf Remark}.  Both Lemmas 1 and 2 have made apparent the speciality of $\alpha=2$ in
the component sums of $C(1,1)$.  When $\alpha=2$, the reduction (10) of $_2F_1$ to 
elementary functions follows.  As regards Lemma 2, analogously to the fact that
the Hurwitz zeta function $\zeta(s,1/2)$ reduces to the Riemann zeta function,
the Lerch zeta function $\Phi(z,2,1/2)$ reduces to the Legendre $\chi_2$ function,
that is in turn a difference of dilogarithms.  It is these reductions that
permit the identification of the full sum $C(1,1)$ as a combination of Clausen 
function values.

\medskip
\centerline{\bf Extensions of harmonic number sums}
\medskip

There are several generalizations of the sums $S(\alpha,\beta)$ of Eq. (4),
of which some we now indicate.  We may introduce
$$S_j(\alpha,\beta) \equiv \sum_{n=0}^\infty {{(-1/\beta)^{3n}} \over {(n+1/\alpha)^j}}
H_n,  ~~~~~~|\beta| > 1.  \eqno(38)$$
For purpose of illustration we consider $j$ to be a positive integer, although
this parameter itself could be extended to nonzero complex numbers $s$.  From
an integral representation of the digamma function we obtain
$$S_j(\alpha,\beta)=\int_0^1\left[\Phi\left(-{t \over \beta^3},j,{1 \over \alpha}
\right) - \Phi\left(-{1 \over \beta^3},j,{1 \over \alpha}\right) \right] {{dt} \over
{t-1}}.  \eqno(39)$$

We may further consider sums for integers $p \geq 1$ and $q \geq 0$
$$S_j(\alpha,\beta,p,q) \equiv \sum_{n=0}^\infty {{(-1/\beta)^{3n}} \over {(n+1/\alpha)^j}}H_{pn+q},  ~~~~~~|\beta| > 1.$$
Extending Eq. (7) we then have
$$S_1(\alpha,\beta,p,q)=\alpha\int_0^1 \left[t^q ~_2F_1\left(1,{1 \over \alpha};1+{1 \over \alpha};-{t^p \over \beta^3}\right)- ~_2F_1\left(1,{1 \over \alpha};1+{1 \over
\alpha};-{1 \over \beta^3}\right)\right]{{dt} \over {t-1}}.$$

As examples relevant to $S(2,2)$ and $C(1,1)$, we consider the sums
$$s(x)\equiv \sum_{n=0}^\infty {{(-x)^{n+1/2}} \over {n+1/2}}H_{2n}, ~~~~~~
t(x)\equiv \sum_{n=0}^\infty {{(-x)^{n+1/2}} \over {n+1/2}}H_{2n+1}, ~~~~|x| \leq 1.$$
We may give a closed form of these sums, and in so doing, provide analytic
continuations to the whole complex plane.  We differentiate and find
$$s'(x)=-\sum_{n=0}^\infty (-x)^{n-1/2}H_{2n}={{-i} \over {(1+x)}}\left[\tan^{-1}
\sqrt{x}-{1 \over {2 \sqrt{x}}}\ln (x+1)\right],$$
and 
$$t'(x)=-\sum_{n=0}^\infty (-x)^{n-1/2}H_{2n+1}={{2\tan^{-1}\sqrt{x}-\sqrt{x}
\ln (x+1)} \over {2ix(1+x)}}.$$
We then integrate both of these equations, noting that $s(0)=t(0)=0$, and find
$$s(x)=-i\tan^{-1}\sqrt{x} ~\ln(x+1),$$
and
$$t(x)=-i[\tan^{-1}\sqrt{x} ~\ln(x+1)+i[\mbox{Li}_2(i\sqrt{x})-\mbox{Li}_2
(-i\sqrt{x})] ].$$
We have then determined the harmonic number sums $s(x)/\sqrt{-x}$ and 
$t(x)/\sqrt{-x}$.  In the latter sum, we may easily write the difference of
dilogarithms as a difference of Clausen function values when $x>0$:
$$-i[\mbox{Li}_2(i\sqrt{x})-\mbox{Li}_2(-i\sqrt{x})]=\omega \ln x+\mbox{Cl}_2
(2\omega)-\mbox{Cl}_2(2\omega+\pi), $$
where $\omega=\tan^{-1} \sqrt{x}$, and we used Eqs. (29) and (30) at $\theta
=\pm \pi/2$.  We may note that when $\sqrt{x}=1/\beta^{3/2}=1/2\sqrt{2}$, 
the angle $\omega$ is none other than that of Proposition 1.

We next introduce generalized harmonic numbers
$$H_n^{(r)} \equiv \sum_{j=1}^n {1 \over j^r}, ~~~~~~~~~~~~H_n \equiv H_n^{(1)}.  \eqno(40)$$
These are given in terms of polygamma functions $\psi^{(j)}$ as
$$H_n^{(r)}={{(-1)^{r-1}} \over {(r-1)!}}\left[\psi^{(r-1)}(n+1)-\psi^{(r-1)}
(1) \right ], \eqno(41)$$
where $\psi^{(r-1)}(1)=(-1)^r(r-1)!\zeta(r)$ and $\zeta$ is the Riemann zeta function.
We now put 
$$S_j(\alpha,\beta,r) \equiv \sum_{n=0}^\infty {{(-1/\beta)^{3n}} \over {(n+1/\alpha)^j}} H_n^{(r)},  ~~~~~~|\beta| > 1.  \eqno(42)$$
By successive differentiation of an integral representation of the digamma function,
the use of Eqs. (40) and (42), and the interchange of summation and integration, we obtain the integral representation
$$S_j(\alpha,\beta,r)=\int_0^1\left[\Phi\left(-{t \over \beta^3},j,{1 \over \alpha}
\right) - \Phi\left(-{1 \over \beta^3},j,{1 \over \alpha}\right) \right]\ln^{r-1} t
{{dt} \over {t-1}}.  \eqno(43)$$

Similarly to our work in the previous section, many other representations of
the sums $S_j(\alpha,\beta)$ and $S_j(\alpha,\beta,r)$ may be constructed.
For instance, in place of Eq. (14) we have the generating function
$$\sum_{n=1}^\infty H_n^{(r)} z^n = {{\mbox{Li}_r(z)} \over {1-z}},  \eqno(44)$$
where Li$_n$ is the polylogarithm function.  Then, for example, we find
$$S_1(\alpha,\beta,r)=\int_0^{-\beta^{-3}} z^{1/\alpha-1} {{\mbox{Li}_r(z)} \over
{1-z}}dz.  \eqno(45)$$ 

\centerline{\bf Series representations of the Clausen function in terms of 
Legendre polynomials}

We present series representations of the function Cl$_2$ in terms of integrals
of the Legendre polynomials $P_n(x)$.  As is well known, these polynomials are
orthogonal on $[-1,1]$.  We have
{\newline \bf Proposition 2}.  
$$\mbox{Cl}_2(\theta)=\left({1 \over 2}-\ln 2\right)\theta+{1 \over 2}\sum_{n=1}^
\infty \left({1 \over n}+{1 \over {n+1}}\right)\int_0^\theta P_n(\cos t) dt.
\eqno(46)$$
Since Catalan's constant $G=\mbox{Cl}_2(\pi/2)$, we have the obvious
{\newline \bf Corollary 1}.  
$$G=\left({1 \over 2}-\ln 2\right){\pi \over 2}+{1 \over 2}\sum_{n=1}^
\infty \left({1 \over n}+{1 \over {n+1}}\right)\int_0^1 {{P_n(u)} \over 
\sqrt{1-u^2}} du.  \eqno(47)$$

After proving Proposition 2 we illustrate it and present related results.

Proof.  By \cite{grad} (p. 1029) or \cite{hansen} (p. 301) we have
$$\sum_{n=1}^\infty {1 \over n} P_n(\cos \theta)=-\ln(\sin \theta/2)
-\ln(1+\sin \theta/2), \eqno(48a)$$
and 
$$\sum_{n=1}^\infty {1 \over {n+1}} P_n(\cos \theta)=-\ln(\sin \theta/2)
+\ln(1+\sin \theta/2)-1. \eqno(48b)$$
We add these two equations and appeal to the integral representation
$$\mbox{Cl}_2(\theta)=-\int_0^\theta \ln\left (2 \sin {t \over 2}\right)dt,
\eqno(49)$$
and the Proposition follows.

{\bf Example}.  Of course Cl$_2(0)=\mbox{Cl}_2(\pi)=0$ and we illustrate how
the latter relation follows from Eq. (46).  Since $P_{2m+1}(x)$ is an odd
function, its integral on $[-1,1]$ is zero and only the $n=2m$ terms 
contribute in Eq. (46), wherein by \cite{grad} (p. 822)
$$\int_{-1}^1 {{P_{2m}(x)} \over \sqrt{1-x^2}}dx=\left[{{\Gamma(m+1/2)} \over
{m!}}\right]^2.  \eqno(50)$$
The sum
$$\sum_{m=1}^\infty\left({1 \over {2m}}+{1 \over {2m+1}}\right)
\left[{{(1/2)_m} \over {(1)_m}}\right]^2=2 \ln 2-1,  \eqno(51)$$
and Cl$_2(\pi)=0$ follows from Proposition 2.

Equation (51) is an instance of the generalized hypergeometric sums
$$\sum_{m=1}\left({1 \over {2m}}+{1 \over {2m+1}}\right)
\left[{{(a)_m} \over {(b)_m}}\right]^2={a^2 \over b^2}\left[{1 \over 2} ~_4F_3
(1,1,a+1,a+1;2,b+1,b+1;1)\right.$$
$$\left. +{1 \over 3} ~_4F_3\left(1,{3 \over 2},a+1,a+1;{5 \over 2},
b+1,b+1;1\right)\right].  \eqno(52)$$
The left side of this equation is easily brought into hypergeometric form
by shifting the summation index and using $(a)_{m+1}=a(a+1)_m$ and $(c)_m/
(c+1)_m=c/(c+m)$.

Related to the example and Corollary 1 we have the separate sums \cite{hansen}
(p. 95)
$${\pi \over 2}\sum_{m=1}^\infty{1 \over {2m}}\left[{{(1/2)_m} \over {(1)_m}}\right]^2
=\pi \ln 2 -2G, \eqno(53a)$$
and \cite{hansen} (p. 92)
$${\pi \over 2}\sum_{m=1}^\infty{1 \over {2m+1}}\left[{{(1/2)_m} \over {(1)_m}}\right]^2 =2G-{\pi \over 2}.  \eqno(53b)$$
Equations (53) are special cases of \cite{grad} (p. 626) or \cite{hansen} (p. 148)
$$\sum_{k=0}^\infty \left[{{(1/2)_k} \over {k!}}\right]^2 {1 \over {ka+b}} x^k
={2 \over {\pi a}}\int x^{b/a-1} {\bf K}(x^{1/2})dx, ~~~~|x| \leq 1, \eqno(54)$$
where $\bf K$ is the complete elliptic integral of the first kind, and
$\int_0^1 {\bf K}(t)dt=2$.  After all, we have \cite{grad} (p. 905)
$${2 \over \pi}{\bf K}(k)= ~_2F_1\left({1 \over 2},{1 \over 2};1;k^2\right)
=\sum_{n=0}^\infty {{(1/2)_n^2} \over {(1)_n^2}} k^{2n}, ~~~~~~|k^2| < 1.
\eqno(55)$$

In consequence of \cite{hansen} (p. 302) or \cite{hold} (p. 284)
$$-\ln(1-x)=1-\ln 2+\sum_{k=1}^\infty {{(2k+1)} \over {k(k+1)}} P_k(x), ~~~~~~
-1 \leq x < 1, \eqno(56)$$
we have
{\newline \bf Proposition 3}
$$\mbox{Li}_2(z)=\int_0^z \left[1-\ln 2+\sum_{k=1}^\infty {{(2k+1)} \over {k(k+1)}} P_k(x)\right]{{dx} \over x}, ~~~~ -1 \leq z < 1, \eqno(57)$$
where the integral may be taken as $\lim_{a \to 0}\int_a^z$.  The singular terms
cancel as $a \to 0$ owing to the contribution of the constant term $P_{2n}(0)$.
The latter constant is given by \cite{grad} (p. 1025)
$$P_{2n}(0)=(-1)^n {{(2n-1)!!} \over {2^n n!}},  \eqno(58)$$
where $(2n-1)!! \equiv (2n-1)(2n-3) \cdots 3 \cdot 1$, and accordingly we have the
sum
$$\sum_{n=1}^\infty {{(4n+1)} \over {2n(2n+1)}} {{(-1)^n} \over \sqrt{\pi}}
{{\Gamma(n+1/2)} \over {n!}}= \ln 2-1.  \eqno(59)$$
In obtaining this equation we used the relation $\Gamma(n+1/2)=\sqrt{\pi}(2n-1)!!/2^n$.
Equation (59) corresponds to a combination of two special cases of the 
hypergeometric summation (6.11.12) of \cite{hansen} (p. 146).
The representation (57) of the dilogarithm function is a parallel of Proposition 2
for the Clausen function. Equation (57) may therefore provide a series representation of the fundamental constant Li$_2(-1)= -\zeta(2)/2$. 

We next have
{\newline \bf Proposition 4}.  Let for $a \in R$
$$\theta(a)=\cos^{-1}\left({{1-a^2} \over {1+a^2}}\right).  \eqno(60)$$
Then 
$$\mbox{Cl}_2(\theta)=2\sum_{k=0}^\infty {a^{k+1} \over {(k+1)}} \int_0^{1/a}
v^{k+1} {{P_k(av)} \over {1+v^2}}dv.  \eqno(61)$$
{\newline \bf Corollary 2}.
$$G=2 \sum_{k=0}^\infty {1 \over {(k+1)}} \int_0^1
v^{k+1} {{P_k(v)} \over {1+v^2}}dv.  \eqno(62)$$

Proof.  From Lemma 2 of \cite{coffey07} we have
$$\mbox{Cl}_2(\theta)=\int_a^\infty \ln\left({{u+a} \over {u-a}}\right){{du} \over
{1+u^2}}.  \eqno(63)$$
From \cite{hansen} (p. 303) we have
$$\sum_{k=0}^\infty {x^{k+1} \over {k+1}} P_k(x)=\tanh^{-1} x ={1 \over 2}
\ln\left({{1+x} \over {1-x}}\right), ~~~~~~|x| <1.  \eqno(64)$$
In order to obtain the Proposition we change variable in Eq. (63) to $v=1/u$ 
and then apply the representation (64).

{\bf Remark}.  The integrals 
$$I_k^{(0)}=\int_0^1 {v^k \over {1+v^2}}P_k(v)dv=\int_0^1 {v^k \over {1+v^2}}
~_2F_1\left[-k,k+1;1;{{1-v} \over 2}\right]dv=a^{(0)}_k+b^{(0)}_k \pi,
\eqno(65)$$
and
$$I_k^{(1)}=\int_0^1 {v^{k+1} \over {1+v^2}}P_k(v)dv=a^{(1)}_k+b^{(1)}_k \ln 2,
\eqno(66)$$
have the stated form, where $a^{(j)}_k$ and $b^{(j)}_k$ are rational numbers.
Similar considerations apply to the family of integrals
$$I_k^{(j)}(a)=\int_0^1 {v^{k+j} \over {a^2+v^2}}P_k(v)dv={1 \over a}\int_0^\infty 
\sin ax \int_0^1 v^{k+j} e^{-x v} P_k(v)dv dx.  \eqno(67)$$

The integrals $I_k^{(j)}(a)$ satisfy the recursion relation
$$I_{k+2}^{(j)} + a^2 I_k^{(j)}=\int_0^1 v^{k+j} P_k(v)dv.  \eqno(68)$$
We have the initial conditions $I_0^{(0)}(1)=\pi/4$, $I_1^{(0)}(1)=1-\pi/4$,
$I_0^{(1)}(1)={1 \over 2}\ln 2$, and $I_1^{(1)}(1)={1 \over 2}(1-\ln 2)$.
Since the type of polynomials integrated on the right side of Eq. (68) can
give only rational numbers, the existence of the recursion relation thus
verifies Eqs. (65) and (66).
In the very special case of $I_k^{(1)}(1)$, we have \cite{grad} (p. 822) 
$\int_0^1 v^{k+1} P_k(v)dv=1/2^{k+1}$.

We have the apparently new integration results contained in
{\newline \bf Lemma 3}.  We have
$$\int_0^1 {v^{2m+1} \over {1+v^2}}P_{2m}(v)dv={1 \over 2^{2m+1}} ~_3F_2\left[
1,m+1,m+{3\over 2};{3 \over 2},2(m+1);-1\right], \eqno(69a)$$
and
$$\int_0^1 {v^{2m+2} \over {1+v^2}}P_{2m+1}(v)dv={1 \over 2^{2(m+1)}} ~_3F_2\left[
1,m+{3 \over 2},m+2;{3 \over 2},2m+3;-1\right]. \eqno(69b)$$

To obtain Eqs. (69) we first write
$$\int_0^1 {v^{k+1} \over {1+v^2}}P_k(v)dv=\sum_{j=0}^\infty (-1)^j\int_0^1 v^{k+2j+1}
P_k(v)dv.  \eqno(70)$$
We then apply \cite{grad} (p. 822) for even and odd indices $k$, and the Lemma
follows.

\medskip
\centerline{\bf Log sine integrals and series and integral representation of the Clausen function}
\medskip

The Log-Sine integrals Ls$_n$ of order $n$ are defined by
$$\mbox{Ls}_n(\theta)=-\int_0^\theta \left(\ln\left|2 \sin {1 \over 2}t\right|
\right)^{n-1}dt.  \eqno(71)$$
In this section we develop the exponential generating function $I(x,u)$ of these
integrals and then use it to write a series and integral representation of the 
Clausen function Cl$_2$.  

We have
\newline{\bf Proposition 5}.  We have for $u>0$
$$\mbox{Cl}_2(u)=2\left[\sin^{-1}\left(\cos{u \over 2}\right)\ln 2-\sum_{j=1}^\infty
{{(1/2)^2_j} \over {(3/2)_j}} \sum_{k=0}^{j-1} {1 \over {(2k+1)}}{{\cos^{2j+1}(u/2)} \over {j!}}\right].  \eqno(72)$$
{\newline \bf Corollary 3}.
$$G=2\left[{\pi \over 4}\ln 2-{1 \over \sqrt{2}}\sum_{j=1}^\infty {{(1/2)^2_j} \over {(3/2)_j}} \sum_{k=0}^{j-1} {1 \over {(2k+1)}}{1 \over {2^j j!}}\right],  \eqno(73)$$
and
$$\mbox{Cl}_2\left({\pi \over 3}\right) ={1 \over 2\sqrt{3}}\left[\psi'\left({1 \over 3}\right)-{2 \over 3}\pi^2\right]= 2\left[{\pi \over 3}\ln 2-{\sqrt{3} \over 2}\sum_{j=1}^\infty {{(1/2)^2_j} \over {(3/2)_j}} \sum_{k=0}^{j-1} {1 \over {(2k+1)}}
{1 \over {j!}}\left({3 \over 4}\right)^j \right].  \eqno(74)$$

Proof.  We put for $2 \pi >u>0$
$$I(x,u) =\int_0^u \exp\left[x\ln\left(2\sin {\theta \over 2}\right)\right]d\theta
=\sum_{n=0}^\infty \int_0^u {x^n \over {n!}} \left[\ln\left(2\sin {\theta \over 2}\right)\right]^n d\theta$$
$$=-\sum_{n=0}^\infty {x^n \over {n!}} \mbox{Ls}_{n+1}(u), \eqno(75)$$
where Ls$_2(\theta)=\mbox{Cl}_2(\theta)$.  Alternatively, we have
$$I(x,u)=\int_0^u \exp\left[\ln\left(2\sin {\theta \over 2}\right)^x\right]d\theta
=2^x \int_0^u \sin^x \left({t \over 2}\right)dt$$
$$=2^x\left[{{\sqrt{\pi} \Gamma\left({{x+1} \over 2}\right)} \over {\Gamma
\left({x \over 2}+1\right)}}-2\cos\left({u \over 2}\right) ~_2F_1\left({1 \over 2},
{{1-x} \over 2};{3 \over 2}; \cos^2 {u \over 2}\right)\right],  \eqno(76)$$
where the first term on the right of Eq. (76) is independent of $u$.

We evaluate
$$\left.{{\partial I(x,u)} \over {\partial x}}\right|_{x=0}=-\mbox{Cl}_2(u).  \eqno(77)$$
Since as $x \to 0$ we have
$$2^x{{\sqrt{\pi} \Gamma\left({{x+1} \over 2}\right)} \over {\Gamma
\left({x \over 2}+1\right)}}=\pi + {\pi^3 \over {24}}x^2 + O(x^3), \eqno(78)$$
the first term on the right side of Eq. (76) does not contribute to the value
Cl$_2(u)$.

For the other term on the right side of Eq. (76) we determine
$${\partial \over {\partial x}} 2^x~_2F_1\left({1 \over 2},{{1-x} \over 2};{3 \over 2}; z \right)=2^x \ln 2~_2F_1\left({1 \over 2},{{1-x} \over 2};{3 \over 2}; z \right)$$
$$+2^x {1 \over 2}\sum_{j=0}^\infty {{(1/2)_j} \over {(3/2)_j}}\left({{1-x} \over 2}
\right)_j\left[\psi\left({{1-x} \over 2}\right)-\psi\left({{1-x} \over 2}+j\right)
\right]{z^j \over {j!}}.  \eqno(79)$$
In order to complete the evaluation of Eq. (77) we use
$$\psi\left(j+{1 \over 2}\right)-\psi\left({1 \over 2}\right)=2\sum_{k=0}^{j-1}
{1 \over {2k+1}}, \eqno(80)$$
and \cite{grad} (p. 51 or 1042)
$$z ~_2F_1\left({1 \over 2},{1 \over 2};{3 \over 2}; z^2 \right)=\sin^{-1} z,  \eqno(81)$$
and the Proposition follows.

From Proposition 5 we have
\newline{\bf Proposition 6}.  We have for $2 \pi > u>0$
$$\mbox{Cl}_2(u)=2\left\{\sin^{-1}\left(\cos{u \over 2}\right)\ln 2-\int_0^1 \left[\sin^{-1}{1 \over \sqrt{2}}\sqrt{1-\sqrt{1-t\cos^2{u \over 2}}}\right. \right.$$
$$\left. \left. -\sqrt{t} \sin^{-1}{1 \over \sqrt{2}}\sqrt{1-\sin{u \over 2}}\right]{{dt} \over{t(t-1)}}\right\}. \eqno(82)$$
{\newline \bf Corollary 4}.
$$G=2\left\{{\pi \over 4}\ln 2-\int_0^1 \left[\sin^{-1}{1 \over \sqrt{2}} \sqrt{1-\sqrt{1-t/2}}-{\pi \over 8}\sqrt{t}\right]{{dt} \over{t(t-1)}}\right\}, \eqno(83)$$
and
$$\mbox{Cl}_2\left({\pi \over 3}\right)=2\left\{{\pi \over 3}\ln 2-\int_0^1 \left[\sin^{-1}{1 \over \sqrt{2}} \sqrt{1-\sqrt{1-{3 \over 4}t}}-{\pi \over 6}\sqrt{t}\right]{{dt} \over{t(t-1)}}\right\}.  \eqno(84)$$

Proof.  In the right side of Eq. (72) we use the integral representation
$$2\sum_{k=0}^{j-1}{1 \over {2k+1}}=\psi\left(j+{1 \over 2}\right)-\psi\left({1 \over
2}\right)=\int_0^1 {{t^{j-1/2}-t^{-1/2}} \over {t-1}}dt,  \eqno(85)$$
with an absolutely convergent integral for integers $j\geq 0$.  We may then
interchange summation and integration, using (cf. Eq. (81))
$$\sum_{j=1}^\infty {{(1/2)^2_j} \over {(3/2)_j}}\left( {{t^{j-1/2}-t^{-1/2}} \over {t-1}}\right){z^j \over {j!}}={2 \over {t(t-1)\sqrt{z}}}\left[\sin^{-1}{1 \over \sqrt{2}}\sqrt{1-\sqrt{1-tz}}\right.$$
$$\left.-\sqrt{t}\sin^{-1} {1 \over \sqrt{2}}\sqrt{1-\sqrt{1-z}} \right].  \eqno(86)$$
We then put $z=\cos^2(u/2)$, take into account the remaining factor of $z$ and the
other term on the right side of Eq. (72), and the Proposition follows.

{\bf Remark}.  Equation (85) that we have applied contains the difference of harmonic
numbers, as we have
$$2\sum_{k=0}^{j-1}{1 \over {2k+1}}=\psi\left(j+{1 \over 2}\right)+\gamma+2\ln 2
=2H_{2j-1}-H_{j-1}.  \eqno(87)$$
This expression also follows by the use of the duplication formula satisfied by
the digamma function,
$$2\psi(2x)=2\ln 2+\psi(x)+\psi\left(x+{1 \over 2}\right).  \eqno(88)$$

\medskip
\centerline{\bf Integrals expressible as Clausen functions}
\medskip

We present integrals that are expressible in terms of the function Cl$_2$.
These integrals are of interest in their own right, and cases of them are
useful in applications.  We also introduce and briefly discuss the very related
and important Lobachevskiy's function $L(x)$.  We have
{\newline \bf Proposition 7}. (a) For $\kappa >0$, and $|u| \leq |\alpha|$ we have
$$\kappa \int_0^u \ln(\sin \kappa x+\sin \alpha)dx=\mbox{Cl}_2(\alpha)-\mbox{Cl}_2
(\kappa u+\alpha)+\mbox{Cl}_2(\alpha-\kappa u+\pi)-\mbox{Cl}_2(\alpha+\pi)-\kappa
u \ln 2,  \eqno(89)$$
and for $A \in R$,
$$\int_0^x \ln|\cos A-\cos kt|dt=-{1 \over k}[\mbox{Cl}_2(kx-A)+\mbox{Cl}_2(kx+A)
+kx\ln 2]. \eqno(90)$$
(b) For $A \in R$, $kx \in R$, $r_1 \equiv \exp(-A)$, $\theta_1 \equiv kx-\pi/2$, and
$$\omega_1 \equiv -\tan^{-1} \left({{r_1 \cos kx} \over {1-r_1 \sin kx}}\right), \eqno(91)$$
we have
$$\int_0^x \ln |\sin kt-\cosh A|dt=(A-\ln 2)x-{1 \over k}\left\{-2A(\omega_1+
\tan^{-1}r_1)+\mbox{Cl}_2(2\tan^{-1}r_1)\right.$$
$$\left. -\mbox{Cl}_2(\pi+2\tan^{-1}r_1)+ \mbox{Cl}_2(2\omega_1) -\mbox{Cl}_2(2\omega_1+2\theta_1)-\mbox{Cl}_2 (2\theta_1)\right \}.  \eqno(92)$$
(c) For $0 \leq b < 2 \pi$ we have
$$\int_0^b {a \over {\sin a}}da=\mbox{Cl}_2(b)-\mbox{Cl}_2(b+\pi)+i{\pi \over 4}
(2b-\pi)+b \ln\left({{1-e^{ib}} \over {1+e^{ib}}}\right)+i {\pi^2 \over 4}, \eqno(93)$$
$$\int_0^b {a \over {\tan a}}da=\mbox{Cl}_2(b)+\mbox{Cl}_2(b+\pi)-i{\pi^2 \over 4}
+b\left[\ln(1-e^{2ib})-i {b \over 2}\right]+{i \over 2}\left(\pi b-b^2+{\pi^2 \over
2}\right), \eqno(94)$$
$$\int_0^b {a^2 \over {\sin^2 a}}da=\mbox{Cl}_2(2b)+ib(\pi-2b)+b[2\ln(1-e^{2ib})
-b \cot b], \eqno(95)$$
and
$$\int_0^b {a^2 \over {\tan^2 a}}da=\mbox{Cl}_2(2b)+ib(\pi-2b)-{1 \over 3}b^3+b[2\ln(1-e^{2ib})-b \cot b]. \eqno(96)$$
(d) Let $0 \leq b < 2 \pi$ and $a \in R$.  Put 
$$\phi=\tan^{-1} (a/\sqrt{1-a^2}), ~~~~~~~~u_{\pm}=ia\pm \sqrt{1-a^2}. \eqno(97)$$
Then 
$$\int_0^b {x \over {\sin x + a}}dx={1 \over \sqrt{1-a^2}}\left[-2\mbox{Cl}_2(\pi+\phi)
+ib\phi+\mbox{Cl}_2(\pi-b+\phi)-\mbox{Cl}_2(\pi-b-\phi)\right.$$
$$\left. +ib\ln\left({{1-e^{-ib}/u_+} \over {1-e^{-ib}/u_-}}\right)\right].  
\eqno(98)$$
Let $0 \leq b < 2 \pi$, $a \in R$, and
$$\phi_a =-\tan^{-1}[2a/(1-a^2)], ~~~~~~~v_+ = \sqrt{{{1+i a} \over {1-i a}}}.
\eqno(99)$$
Then 
$$\int_0^b {x \over {\tan x + a}}dx={1 \over {8v_+}}\left\{{2 \over {1+a^2}}\left[
\mbox{Cl}_2(\phi_a)-\mbox{Cl}_2(\phi_a-2b)+2b\ln(1-e^{i(\phi_a-2b)}) \right] \right.$$
$$\left. {{2ib} \over {(1+ia)}}\left[b+ {{b-\phi_a+\pi} \over {1-ia}}\right]
\right \}.  \eqno(100)$$
In Eqs. (97 and (99), the value of $\tan^{-1}$ is chosen so that it lies in
the range $[0,2\pi]$.

From Proposition 7(a) follows
{\newline \bf Corollary 5}.  For $|u| \leq |\alpha|$ we have
$$\int_0^u \ln(\sin^2 x-\sin^2 \alpha)dx={1 \over 2}\left[\mbox{Cl}_2(2(\alpha-u))
-\mbox{Cl}_2(2(\alpha+u))\right]-2u\ln 2.  \eqno(101)$$
From Proposition 7(c) follows
{\newline \bf Corollary 6}.  Integrals of the form
$$\int_0^b {{x dx} \over {\sin(x+a)}} \eqno(102)$$
are expressible in terms of differences of Clausen function values and logarithms.

Proof.  For part (a) we write
$$\kappa \int_0^u \ln(\sin \kappa x+\sin \alpha)dx=\int_0^u \ln \left[2\sin{1 \over 2}
(\kappa x+\alpha) \cos{1 \over 2}(\kappa x-\alpha)\right]dx$$
$$={1 \over \kappa}\int_\alpha^{\kappa u+\alpha} \ln\left(2 \sin {\theta \over 2}
\right)d\theta+{2 \over \kappa}\int_{-\alpha/2}^{(\kappa u-\alpha)/2} \ln(\cos
\phi) d\phi.  \eqno(103)$$
We change variable in the latter integral, and recall from Eq. (1) that
$$-\int_0^\theta \ln\left|\sin{t \over 2}\right|dt=\mbox{Cl}_2(\theta)+\theta \ln 2,
\eqno(104)$$
yielding Eq. (89).  
For Eq. (90), we proceed similarly, writing
$$\int_0^x \ln|\cos A-\cos kt|dt=\int_0^x \ln\left|2\sin\left({{kt+A} \over 2}
\right)\sin\left({{kt-A} \over 2}\right)\right|dt.  \eqno(105)$$
We then change variables in the integrals, again use Eq. (104), and Eq. (90) follows.

For part (b), we begin with (\cite{herbert}, Eq. (I-10))
$$\int_0^x \ln |\sin kt-\cosh A|dt=(A-\ln 2)x+{2 \over k}\sum_{n=1}^\infty {e^{-nA}
\over n^2} \left[\sin n\left({\pi \over 2}-kx\right)-\sin {{n\pi} \over 2}\right].  \eqno(106)$$
We rewrite this equation in terms of differences of dilogarithms:
$$\int_0^x \ln |\sin kt-\cosh A|dt=(A-\ln 2)x+{i \over k}\left[\mbox{Li}_2(ie^{-A})
-\mbox{Li}_2(-ie^{-A})\right.$$
$$\left.+\mbox{Li}_2(-ie^{-A+ikx})-\mbox{Li}_2(ie^{-A-ikx})\right],
\eqno(107)$$
where we recognize the first difference of dilogarithms on the right side of this equation as $2\chi_2(ie^{-A})$.  

Both pairs of differences of dilogarithms in Eq. (107) are pure imaginary.  For
we may note that for $r \geq 0$, 
$$\mbox{Li}_2(ir)-\mbox{Li}_2(-ir)=\sum_{n=1}^\infty {r^n \over n^2}i^n[1-(-1)^n],
\eqno(108)$$ 
and for $kx$ real 
$$\mbox{Li}_2(-ire^{ikx})-\mbox{Li}_2(ire^{-ikx})=\sum_{n=1}^\infty {r^n \over n^2}i^n[(-1)^ne^{ikxn}-e^{-ikxn}]$$  
$$=2\sum_{n=1}^\infty {r^n \over n^2}i^n \left\{\begin{array}{c} 
~~ i \sin kxn ~~~(n ~~\mbox{even})\\ - \cos kxn ~~~(n ~~\mbox{odd}) \end{array}\right\}.
\eqno(109)$$
Whether $n$ is even or odd in these sums the result is then pure imaginary.

By Eqs. (29) and (30) we find
$$\mbox{Im}~\left[\mbox{Li}_2(ie^{-A})-\mbox{Li}_2(-ie^{-A})\right]=-2A\tan^{-1}
(e^{-A})+\mbox{Cl}_2(2\tan^{-1}r_1)-\mbox{Cl}_2(\pi+2\tan^{-1}r_1).  \eqno(110)$$
Similarly applying Eqs. (29) and (30) to the second difference of dilogarithms
on the right side of Eq. (107) we determine
$$\mbox{Im}~\left[\mbox{Li}_2\left(e^{-A}e^{-i{\pi \over 2}+ikx}\right)-
\mbox{Li}_2\left(e^{-A}e^{i{\pi \over 2}-ikx}\right)\right]=2\omega_1 \ln r_1
+\mbox{Cl}_2(2\omega_1)-\mbox{Cl}_2(2\omega_1+2\theta_1)-\mbox{Cl}_2
(2\theta_1).  \eqno(111)$$
Combining Eqs. (102), (110), and (111) gives part (b) of the Proposition.

For part (c), by a change of variable we have
$$\int_0^b {a \over {\sin a}}da=-i\int_1^{e^{ib}}\left[{1 \over {u-1}}-{1 \over
{u+1}}\right]\ln u ~du$$
$$=-i\left[-\mbox{Li}_2(1-c)-{\pi^2 \over {12}}-\ln c\ln(c+1)-\mbox{Li}_2(-c)
\right], \eqno(112)$$
where $c \equiv \exp(ib)$.  We then use the relation (e.g. \cite{sri}, p. 107)
$$\mbox{Li}_2(1-z)=-\mbox{Li}_2(z)+{\pi^2 \over 6}-\ln z\ln(1-z). \eqno(113)$$  
We also twice use
the relation (e.g., \cite{sri}, p. 111)
$$\mbox{Li}_2(e^{ib})={\pi^2 \over 6}-{b \over 4}(2\pi-b)+i \mbox{Cl}_2(b), ~~~~~~
0 \leq b \leq 2 \pi. \eqno(114)$$
Combining terms, we obtain Eq. (93).  

Equation (94) follows similarly, starting with
$$\int_0^b {a \over {\tan a}}da=-{i\over 2}\int_1^{e^{ib}}\left[{1 \over {u-1}}
-{1 \over {u+1}}\right]\left(u + {1 \over u}\right)\ln u ~du $$
$$=-{i \over 2}\left\{{\pi^2 \over 6}-\ln c[\ln c-2\ln(c+1)]+2[\mbox{Li}_2(-c)
-\mbox{Li}_2(1-c)]\right\}. \eqno(115)$$
Again using Eqs. (113) and (114) leads to Eq. (94).

For Eq. (95) we may write 
$$\int_0^b {a^2 \over {\sin^2 a}}da=-4i\int_0^{e^{ib}} {{u\ln^2 u ~du} \over
{(u^2-1)^2}}$$
$$=-i\int_1^{e^{ib}}\left[{1 \over {(u-1)^2}}-{1 \over {(u+1)^2}}\right] 
\ln^2 u ~du. \eqno(116)$$
The necessary integrals may be found in Section 3.12 of Ref. \cite{duke},
expressing the result in terms of a dilogarithm.  Then the use of Eq. (114)
gives Eq. (95).
For Eq. (96) we simply use the identity $\cot^2 x=\csc^2 x-1$ and Eq. (95).

For part (d), by changes of variable we have
$$\int_0^b {x \over {\sin x +a}}dx={{2i} \over {(u_--u_+)}}\int_1^{e^{-ib}}\left[
{1 \over {u-u_-}}-{1 \over {u-u_+}}\right]\ln u ~du $$
$$={{2i} \over {(u_--u_+)}}\left[-\mbox{Li}_2\left({1 \over u_+}\right)-
\mbox{Li}_2\left({1 \over u_-}\right)+\mbox{Li}_2\left({e^{-ib} \over u_-}\right)-
\mbox{Li}_2\left({e^{-ib} \over u_+}\right)+ib\ln\left({{1-e^{-ib}/u_+}
\over {1-e^{-ib}/u_-}}\right)\right]$$
$$={{2i} \over {(u_--u_+)}}\left[-\mbox{Li}_2\left(-u_-\right)-
\mbox{Li}_2\left(-u_+\right)+\mbox{Li}_2\left(-e^{-ib}u_+\right)-
\mbox{Li}_2\left(-e^{-ib} u_-\right)+ib\ln\left({{1-e^{-ib}/u_+}
\over {1-e^{-ib}/u_-}}\right)\right]. \eqno(117)$$
We then repeatedly apply Eq. (110), yielding Eq. (98). 

For Eq. (100), we have, with $v=\exp(-ix)$,
$$\int_0^b {x \over {\tan x +a}}dx={i \over {(1-ia)}}{1 \over v_+}\int_1^{e^{-ib}}
{{\ln v (v+v^{-1}) dv} \over {(v-v_+)(v+v_+)}}$$
$$={i \over {8v_+}}{1 \over {(1+ia)}}\left\{-2ib[ib+(1+v_+^2)\ln(1-e^{-2ib}/v_+^2)]
\right.$$
$$+\left.(1+v_+^2)\left[\mbox{Li}_2\left({e^{-2ib} \over v_+^2}\right) - 
\mbox{Li}_2\left({1 \over v_+^2}\right)\right] \right\}. \eqno(118)$$
We now use Eq. (114) to write
$$\mbox{Li}_2\left({e^{-2ib} \over v_+^2}\right)-\mbox{Li}_2\left({1 \over v_+^2}\right)=\mbox{Li}_2[e^{i(\phi_a-2b)}] - \mbox{Li}_2(e^{i\phi_a})
=i[\mbox{Cl}_2(\phi_a-2b)-\mbox{Cl}_2(\phi_a)]+b(b-\phi_a+\pi). \eqno(119)$$
Substituting this equation into Eq. (118), applying the definition of $v_+^2$,
and rearranging gives Eq. (100), completing the Proposition.  

Corollary 5 is found by adding Eq. (89) at $\kappa=1$ with Eq. (89) at $\kappa=1$
and $\alpha \to -\alpha$.  Then the duplication formula (37) is used.  
The result, Eq. (97), is equivalent to (\cite{grad}, 4.226.5, p. 528).

Corollary 6 follows readily by a simple change of variable and the use of an
elementary integral:
$$\int_0^b {{x dx} \over {\sin(x+a)}}=\int_a^{a+b} {{y dy} \over {\sin y}}
-a\left[\ln \tan\left({{a+b} \over 2}\right)-\ln \tan\left({a \over 2}\right)
\right]. \eqno(120)$$
The remaining integral on the right side of this equation may be evaluated by
using Eq. (93).  

{\bf Remarks}.  Equation (90) of Proposition 7(a) may also be determined by writing
Eq. (I-6) of Ref. \cite{herbert} as a pair of differences of dilogarithms.

Obviously Proposition 7(c) encompasses the very special cases
$$\int_0^{\pi/2}{a \over {\sin a}}da=2G, ~~~~~~~~
\int_0^{\pi/2}{a \over {\tan a}}da={\pi \over 2}\ln 2, \eqno(121)$$
and
$$\int_0^{\pi/4}{a^2 \over {\sin^2 a}}da=G-{\pi \over {16}}(\pi-4\ln 2),$$
$$\int_0^{\pi/4}{a^2 \over {\tan^2 a}}da=G-{\pi \over {16}}(\pi-4\ln 2)-{\pi^3
\over {192}}. \eqno(122)$$

The results of Proposition 7(c) are not new (e.g., \cite{lewin}).  However, the
method of proof may be.
This part of Proposition 7 may be readily extended to integrals of the form
$$\int_0^b{a^k \over {\sin a}}da, ~~~~~~~~ \int_0^b{a^k \over {\tan a}}da, \eqno(123)$$
and
$$\int_0^b{a^k \over {\sin^k a}}da, ~~~~~~~~ \int_0^b{a^k \over {\tan^k a}}da. 
\eqno(124)$$
The results are expressible in terms of Li$_{k+1}$ and lower order polylogarithms.

It is perhaps worth stressing that when $b$ in Proposition 7(c) is a 
rational multiple of $\pi$, the result may be expressed in terms of values
of the trigamma, sine, and logarithm functions.  In addition, Proposition 7(c)
allows us to relate sums over Bernoulli numbers $B_j$ to Clausen and logarithm
function values.  From termwise integration of known expansions for the csc and
cot functions (e.g., p. 35 of \cite{grad}), we have
$$\int_0^b{a \over {\sin a}}da=b+\sum_{k=1}^\infty {{2(2^{2k-1}-1)|B_{2k}|} \over
{(2k+1)!}} b^{2k+1}, ~~~~~~~~ b^2 < \pi^2, \eqno(125)$$
and
$$\int_0^b{a \over {\tan a}}da=b-\sum_{k=1}^\infty {{2^{2k}|B_{2k}|} \over
{(2k+1)!}} b^{2k+1}, ~~~~~~~~ b^2 < \pi^2. \eqno(126)$$
In these equations, $|B_{2k}|=2(2k)!\zeta(2k)/(2\pi)^{2k}$.  Therefore, we
have, for instance,
$$\int_0^b{a \over {\tan a}}da=b-\sum_{k=1}^\infty {1 \over {(2k+1)}}{{\zeta(2k)}
\over {(2\pi)^{2k}}} (2b)^{2k+1}, \eqno(127)$$
while (e.g., \cite{hansen}, p. 356)
$$2\sum_{k=1}^\infty {1 \over {(2k+1)}}{{\zeta(2k)} \over {(2\pi)^{2k}}} t^{2k+1}
=t-t\ln\left(2 \sin {t \over 2}\right)- \mbox{Cl}_2(t). \eqno(128)$$
Therefore, we obtain
$$\int_0^b{a \over {\tan a}}da=b \ln(2 \sin b)+{1 \over 2}\mbox{Cl}_2(2b). \eqno(129)$$
This equation is in agreement with Eq. (94) due to the duplication formula (37).

The standard reference \cite{grad} does not employ the Clausen function.
However, the Lobachevskiy's function $L$ does appear.  This function, given by
$$L(x)=-\int_0^x \ln |\cos t|dt, \eqno(130)$$
is directly related to Cl$_2$ as
$$L\left({\pi \over 2}\pm\theta\right)=\left({\pi \over 2}\pm\theta\right)\ln 2
\pm {1 \over 2}\mbox{Cl}_2(2\theta), \eqno(131a)$$
and
$$L(x)={1 \over 4}[\mbox{Cl}_2(4x)-2\mbox{Cl}_2(2x)]+x\ln 2.  \eqno(131b)$$
Therefore, all of the various results written in terms of $L$ in Ref. \cite{grad}
may be re-expressed in terms of the Clausen function.  Entries containing $L$
or its equivalent appear on such pages as 347, 354, 355, 526-531, 557, 591-593, 598, 
and 933.  As an example, we have \cite{grad} (p. 355)
$$\int_0^\infty {{x \cosh x} \over {\cosh 2x-\cos 2t}}dx=\mbox{csc}~ t\left[-{1 \over 4} \mbox{Cl}_2(2t)+\mbox{Cl}_2(t)\right].  \eqno(132)$$
Of course, Lobachevskiy's function is very important in finding volumes of
polyhedra in hyperbolic space (e.g., \cite{coxeter,herbert,milnor}).
A result more general than Eq. (132) is proved in Appendix D.  

In Proposition 7(b) we also see the appearance of Lobachevskiy's angle of
parallelism $\Pi_L$.  This function is given by $\Pi_L(x)=2 \tan^{-1} e^{-x}$ for
$x \geq 0$, and $\Pi_L(x)=\pi-\Pi_L(-x)$ for $x<0$ \cite{grad} (p. 43).

Values of the Lobachevskiy's function have appeared in models in statistical
mechanics and quantum field theory, among other applications.  In particular, the
value $L(\pi/6)$ is needed.  We evaluate this as
$$L\left({\pi \over 6}\right)={\pi \over 6}\ln 2-{1 \over 2}\sum_{k=1}^\infty
{{\sin \pi k/3} \over k^2} = {\pi \over 6}\ln 2-{1 \over {12\sqrt{3}}}\left[\psi'
\left({1 \over 3}\right)-\psi'\left({2 \over 3}\right)\right]$$
$$={\pi \over 6}\ln 2-{1 \over {6\sqrt{3}}}\left[\psi'\left({1 \over 3}\right)- {2 
\over 3}\pi^2\right], \eqno(133)$$
wherein we used the reflection formula for the trigamma function $\psi'$.  This
value provides the ground-state entropy per spin for the triangular 
antiferromagnet \cite{wannier} (p. 364), \cite{choy} (p. 7363)
$$S/k_B={1 \over 2}\ln 2-{3 \over \pi}L\left({\pi \over 6}\right)
={1 \over {3\sqrt{3}}}\left[{3\over {2\pi}}\psi'\left({1 \over 3}\right)-\pi\right],
\eqno(134)$$
where $k_B$ is Boltzmann's constant, as well as a contribution to the curvature of an effective potential (\cite{arnold}, Eqs. (6.31) and (B8)), $(\pi/6)\ln 2-L(\pi/6)$.
It also appears in an approximate formula for the two-loop Higgs mass correction \cite{espinosa} (Eq. (A.29)).
The value $\pi S/k_B$ of Eq. (134) gives the volume of a regular tetrahedron of
6 edges \cite{coxeter} (pp. 28-29).

It is worth noting that from Eqs. (131b) and (37) we have the relation
$${\pi \over 6}\ln 2-L\left({\pi \over 6}\right)={1 \over 3}\mbox{Cl}_2\left({\pi 
\over 3}\right).  \eqno(135)$$

The following Proposition is a generalization of Lemma 2(a) of Ref. \cite{coffey07}.
\newline{\bf Proposition 8}.  Let $a$ and $b$ be real numbers and set
$$\theta =\cos^{-1}\left({{1-a^2} \over {1+a^2}}\right),  \eqno(136)$$
$$r \equiv {{b+a} \over {b-a}}, ~~~~\tan \omega={{r\sin \theta} \over {1-r\cos \theta}}, ~~~~\chi = \pi-\theta-\omega.  \eqno(137)$$
Then we have 
$$\int_b^\infty \ln\left({{u+a} \over {u-a}}\right){{du} \over {1+u^2}}=
\mbox{Cl}_2(\pi-\theta) -{1 \over 2}[\mbox{Cl}_2(2\omega) +\mbox{Cl}_2(2\chi)].
\eqno(138)$$

Proof.  We consider the integrals
$$\int_b^\infty \ln^n\left({{u+a} \over {u-a}}\right){{du} \over {1+u^2}}  
={{2a} \over {(1+a^2)}}\int_1^r {{\ln^n v ~dv} \over {v^2+2\left({{a^2-1}
\over {a^2+1}}\right)v+1}}, \eqno(139)$$
where we changed variable to $v=(u+a)/(u-a)$.  We then apply the integral
representation 
$${1 \over 2}[\mbox{Cl}_2(2\theta)+\mbox{Cl}_2(2\omega)+\mbox{Cl}_2(2\chi)]
=-\sin \theta \int_0^r {{\ln y ~dy} \over {1-2 y \cos \theta +y^2}},  \eqno(140)$$
with $\theta$, $\omega$, and $\chi$ as given in Eqs. (136) and (137), at $n=1$, obtaining
$$\int_b^\infty \ln\left({{u+a} \over {u-a}}\right){{du} \over {1+u^2}}  
=-{1 \over 2}[\mbox{Cl}_2(2\theta)+\mbox{Cl}_2(2\omega)+\mbox{Cl}_2(2\chi)]
+\mbox{Cl}_2(\theta).  \eqno(141)$$
Lastly, we use the duplication formula (37).

The representation (140) may be obtained from the elementary sum
$$\sum_{n=1}^\infty {{\cos 2nx} \over n}y^n=-{1 \over 2}\ln(1-2y \cos 2x+y^2), 
~~~~~~|y| \leq 1, \eqno(142)$$
according to the following steps.  We differentiate with respect to $x$, 
rearrange, multiplying both sides by $\ln y$, and integrate with respect to
$y$, gaining, after putting $x \to x/2$,
$$\sum_{n=1}^\infty {z^n \over n^2}(n\ln z-1)=\sin x\int_0^z {{\ln y ~dy} \over
{1-2y\cos x+y^2}}. \eqno(143)$$
Writing the left side in terms of Clausen functions leads to Eq. (140).

The result (138) may also be verified by differentiation with respect to $b$.

{\bf Remark}.  The integral of Eq. (138) may be equivalently written as
$$\int_b^\infty \ln\left({{u+a} \over {u-a}}\right){{du} \over {1+u^2}}  
=\int_{\tan^{-1} b}^{\pi/2} \ln \left({{\tan \phi+a} \over {\tan \phi-a}}\right)
d\phi.  \eqno(144)$$
In this form, Proposition 8 subsumes the special case at $b=0$ and $a=1$ effectively
given on p. 530 of \cite{grad}.  

\medskip
\centerline{\bf A $_3F_2$ function as Cl$_2$}
\medskip

We have
\newline{\bf Proposition 9}.
$$_3F_2\left({1 \over 2},{1 \over 2},{1 \over 2};{3 \over 2},{3 \over 2};z\right)
={1 \over {2\sqrt{z}}}\left[\mbox{Cl}_2(2\sin^{-1}\sqrt{z})+2\sin^{-1}\sqrt{z} ~
\ln 2+{{\ln z} \over 2}\left(\pi-2\sin^{-1}\sqrt{1-z}~\right)\right].  \eqno(145)$$

{\bf Corollary 7}.  
$$G=\sqrt{2}_3F_2\left({1 \over 2},{1 \over 2},{1 \over 2};{3 \over 2},{3 \over 2};
{1 \over 2} \right)-{\pi \over 4}\ln 2.  \eqno(146)$$

{\bf Corollary 8}.  Let
$$\Psi(x) \equiv \int_0^x {{\sin^{-1} t} \over t} dt = \int_0^{\sin^{-1} x}
t \cot t ~dt.$$
Then we have
$$\Psi(x)={1 \over 2}\left[\mbox{Cl}_2(2\sin^{-1}x)+2\sin^{-1}x ~
\ln 2+\left(\pi-2\sin^{-1}\sqrt{1-x^2}~\right)\ln x\right].$$

Proof.  We first note from Eq. (1) that
$$\mbox{Cl}_2(\theta)=-2\left[{\theta \over 2}\ln 2+\int_0^{\theta/2} \ln \sin \phi
~d \phi\right].  \eqno(147)$$
We then change variable to $e^{-x}=\sin \phi$, giving
$$\mbox{Cl}_2(\theta)=-\theta \ln 2+2\int_{-\ln(\sin \theta/2)}^\infty {{x dx}\over
\sqrt{e^{2x}-1}}.  \eqno(148)$$

On the other hand, we have
$$_3F_2\left({1 \over 2},{1 \over 2},{1 \over 2};{3 \over 2},{3 \over 2};z\right)
=\sum_{n=0}^\infty {{\Gamma(n+1/2)} \over {\sqrt{\pi} n!}}{z^n \over {(2n+1)^2}}.
\eqno(149)$$
Then upon interchange of summation and integration, justified on the basis of
absolute convergence, we have
$$_3F_2\left({1 \over 2},{1 \over 2},{1 \over 2};{3 \over 2},{3 \over 2};z\right)
=\sum_{n=0}^\infty {{(1/2)_n} \over {(1)_n}} z^n \int_0^\infty e^{-(2n+1)x} x dx
=\int_0^\infty {{x dx} \over \sqrt{e^{2x}-z}}.  \eqno(150)$$
We now write $z^{-1}=\exp(-\ln z)$ and change variable to $y=x-(\ln z)/2$, giving
$$_3F_2\left({1 \over 2},{1 \over 2},{1 \over 2};{3 \over 2},{3 \over 2};z\right)
={1 \over \sqrt{z}}\int_{-(\ln z)/2}^\infty {{\left(y+{1 \over 2}\ln z\right)dy} 
\over \sqrt{e^{2y}-1}}.  \eqno(151)$$
We then evaluate the second elementary integral on the right side of this equation.
For the first integral on the right side of Eq. (151) we compare with the
relation (148).  We accordingly put $\sqrt{z}=\sin \theta/2$ and the Proposition
follows.  

Corollary 7 obviously obtains for the value Cl$_2(\pi/2)$.  Another very special
case of Proposition 9 is
$$_3F_2\left({1 \over 2},{1 \over 2},{1 \over 2};{3 \over 2},{3 \over 2};1 \right)
={\pi \over 2}\ln 2=L\left({\pi \over 2}\right). \eqno(152)$$

Corollary 8 follows from Proposition 9 by noting that
$$\Psi(x)= x~_3F_2\left({1\over 2},{1\over 2},{1\over 2};{3\over 2},{3\over 2};x^2
\right).$$

In particular, the special values of $\Psi(1/3)$ and $\Psi(4\sqrt{2}/9)$ in terms of
Cl$_2$ may be observed.

The relation of a certain $_{k+1}F_k$ function to a binomial sum of Log-sine
integrals is shown in the following extension.

{\bf Lemma 3}.  We have for integers $k\geq 1$
$$_{k+1}F_k\left({1 \over 2},{1 \over 2},\ldots,{1 \over 2};{3 \over 2},\ldots,
{3 \over 2};z\right)$$
$$={1 \over {(k-1)!}}{1 \over \sqrt{z}}\sum_{\ell=0}^{k-1}
(-1)^{\ell}{{k-1} \choose \ell}\left({1 \over 2} \ln z\right)^{k-\ell-1}
\int_0^{\sin^{-1}\sqrt{z}} \ln^\ell \sin \phi ~d\phi.  \eqno(153)$$

Proof.  We have
$$_{k+1}F_k\left({1 \over 2},{1 \over 2},\ldots,{1 \over 2};{3 \over 2},\ldots,
{3 \over 2};z\right)=\sum_{n=0}^\infty {z^n \over {n!}} {{(1/2)_n} \over {(2n+1)^k}}$$
$$={1 \over {(k-1)!}}\sum_{n=0}^\infty {z^n \over {n!}} (1/2)_n\int_0^\infty
e^{-(2n+1)x} x^{k-1} dx.  \eqno(154)$$
Interchanging summation and integration we find
$$_{k+1}F_k\left({1 \over 2},{1 \over 2},\ldots,{1 \over 2};{3 \over 2},\ldots,
{3 \over 2};z\right)={1 \over {(k-1)!}}\int_0^\infty  {{x^{k-1} dx}  \over \sqrt{
e^{2x}-z}}$$
$$={1 \over {(k-1)!}}{1 \over \sqrt{z}}\int_{-(\ln z)/2}^\infty
{{\left(y+{1 \over 2} \ln z\right)^{k-1}dy} \over \sqrt{e^{2y}-1}}.  \eqno(155)$$
Binomial expansion of the numerator of the integrand and the change of variable
$y=-\ln(\sin \phi)$ then completes the Lemma.


\medskip
\centerline{\bf Log trigonometric integrals, three-electron integrals, and $C(1,1)$}
\medskip

Rajantie \cite{raj} has derived the integral expression
$${{C(1,1)}\over 2^{5/2}}=\int_0^1 \left(\ln {3 \over 4}+\ln{{x+3} \over {x+2}}
+{x^2 \over {x^2-4}}\ln {4 \over {x+2}}+{x \over {x+2}}\ln {{x+3} \over 3}\right)
{{dx} \over \sqrt{3-x^2}}.  \eqno(156)$$
By simple rearrangement and partial fractions, this expression takes the form
$${{C(1,1)}\over 2^{5/2}}=\int_0^1 \left[\left({1 \over {x-2}}-{1 \over {x+2}}\right)
\ln 4+2\left(1-{1 \over {x+2}}\right)\ln(x+3)\right.$$
$$\left. -\left(2+{1 \over {x-2}}+{1 \over {x+2}}
\right)\ln(x+2)\right]{{dx} \over \sqrt{3-x^2}},  \eqno(157)$$
where the elementary contribution is given by
$$\ln 4\int_0^1 \left({1 \over {x-2}}-{1 \over {x+2}}\right){{dx} \over \sqrt{3-x^2}}  
=\ln 4\left[\tan^{-1}\left({1 \over \sqrt{2}}\right)-\tan^{-1}\left({5 \over \sqrt{2}}
\right)\right]=\theta_+ \ln 4. \eqno(158)$$
We recall that the angle $\theta_+$ is defined just above Eq. (4), and therefore that
Eq. (158) provides precisely the non-Clausen portion of $S(2,2)$.

We have 
\newline{\bf Corollary 9}.  Integrals of the form
$$J(c,d)=\int_0^1 {{\ln(x+c)} \over \sqrt{d^2-x^2}}dx, ~~~~~d >1,  ~~~~ c/d > 1, 
\eqno(159)$$
may be evaluated in terms of the Clausen function Cl$_2$ and logarithms.

Proof.  By an easy change of variable, we have
$$J(c,d)=\ln d \sin^{-1}\left({1 \over d}\right)+\int_0^{\sin^{-1} (1/d)}\ln(\sin \theta + c/d)d\theta.  \eqno(160)$$
We then apply Proposition 7(b) at $x=\sin^{-1} (1/d)$, $k=-1$, and $\cosh A=c/d$.
This gives $A=\ln(c/d+\sqrt{c^2/d^2-1})$, and the Corollary is complete.

{\bf Remark} and emphasis.  For the cases of interest from Eq. (157) we have
$c=2$ and $3$ and $d=\sqrt{3}$, giving respectively the values $A={1 \over 2}
\ln{3}$ and $\ln(\sqrt{2}+\sqrt{3})$, $r_1=\sqrt{3}$ and $\sqrt{2}+\sqrt{3}$,
$\tan^{-1} r_1=\pi/3$ and $\tan^{-1}(\sqrt{2}+\sqrt{3})$, and
$\omega_1=-\tan^{-1}(1/\sqrt{2})$ and $-\tan^{-1}[(2\sqrt{2}+\sqrt{3})/5]$.

It turns out that not only have three-electron atomic integrals been well studied,
but these have a momentum space representation that is specifically of interest
for three-loop Feynman diagrams.  We have
\newline{\bf Proposition 10}.  The conjectured relation (2) holds, or, equivalently, 
we have 
$$C^{\mbox{Tet}}=2\sqrt{2}\left[-3\mbox{Cl}_2(2\theta_+)+6\mbox{Cl}_2\left(\pi+2\tan^{-1}\left({1 \over \sqrt{2}}\right)\right)+2\mbox{Cl}_2\left((\pi-2\tan^{-1}\left({5 \over \sqrt{2}}\right)\right)\right].  \eqno(161)$$

Proof.  Define, as in Eq. (1) of Ref. \cite{harris}, the generating integral
$$I(\alpha_1,\alpha_2,\alpha_3,\alpha_{12},\alpha_{23},\alpha_{31}) \equiv
\int {{\exp(-\alpha_1r_1-\alpha_2r_2-\alpha_3r_3-\alpha_{12}r_{12}-\alpha_{23}
r_{23}-\alpha_{31}r_{31})} \over {r_1r_2r_3r_{12}r_{23}r_{31}}}$$
$$~~~~~~~~~~~~~~~~~~~~~~~~~~~~~~~~\times d^3r_1d^3r_2d^3r_3, 
\eqno(162)$$
where Re $\alpha_i$, $\alpha_{ij} \geq 0$, at least one of $\alpha_i$, $\alpha_{ij}$
is nonzero, $r_{ij}\equiv |\vec{r}_i-\vec{r}_j|$, and each integration is over $R^3$.  Comparing to Eq. (1) of Ref. \cite{broadhurst}, we find that
$$C^{\mbox{Tet}}={1 \over {8\pi^3}}I(1,1,1,1,1,1) \equiv {1 \over {8\pi^3}}
\mbox{SRP}, \eqno(163)$$
wherein SRP denotes "standard reference point" \cite{fromm,harris}.  
This relation follows due to the Fourier representation
$${e^{-\alpha r} \over r}={1 \over {(2\pi)^3}}\int {{4\pi} \over {p^2+\alpha^2}}
e^{i\vec{p}\cdot \vec{r}} d^3p.  \eqno(164)$$
Put $\sigma=i\sqrt{2}$, according to Eq. (6) of Ref. \cite{harris} when
all parameters in Eq. (162) are unity.  Then Eq. (39) of this reference applies,
$$I={{16\pi^3} \over {|\sigma|}}\left[-2\sum_{j=1}^3 \mbox{Cl}_2\left(\pi-2\tan^{-1}
\left({\Gamma_j \over {|\sigma|}}\right)\right)+\sum_{j=0}^3\sum_{k=0}^3 \mbox{Cl}_2\left(\pi-2\tan^{-1}\left({{\gamma_k^{(j)}} \over {|\sigma|}}\right) 
\right ) \right], \eqno(165)$$
where $\sigma=i|\sigma|$.
In this equation, per the special symmetry of the SRP and Eqs. (8), (9), and (21)
of Ref. \cite{harris}, we have $\Gamma_j=-7/4$, $j=1,2,3$, $\gamma_j^{(j)}=5$,
$j=0,1,2,3$, and $\gamma_k^{(j)}=-1$ for $k\neq j$ and $k,j=0,1,2,3$.
The use of the relation $\tan^{-1}x+\tan^{-1}(1/x)=\pi/2$ for $x>0$ and the $2\pi$
periodicity of Cl$_2$ with Eq. (165) gives Eq. (161).

{\bf Remarks}.  Both of the works \cite{fromm,harris} use the odd function of $z$
$$v(z) \equiv {1 \over 2}\left[\mbox{Li}_2\left({{1-z} \over 2}\right)-
\mbox{Li}_2\left({{1+z} \over 2}\right)\right]+{1\over 4}\left[\ln^2\left({{1-z}
\over 2}\right)-\ln^2\left({{1+z} \over 2}\right)\right].  \eqno(166)$$
For numerical purposes, Harris \cite{harris} introduces the function
$$\bar{v}(z)=v(z)+2z\ln 2, \eqno(167a)$$
and the expansion
$$\bar{v}(z)=\sum_{n=1}^\infty C_n z^{2n+1}, \eqno(167b)$$
where
$$C_n \equiv -{2 \over {2n+1}}\ln 2 - D_n.  \eqno(167c)$$
We note that the coefficients $D_n$ are expressible in terms of harmonic
and generalized harmonic numbers, together with binomial coefficients.  This
follows from the expansions
$${1\over 4}\left[\ln^2\left({{1-z} \over 2}\right)-\ln^2\left({{1+z} \over 2}\right)
\right]=\sum_{n=1}^\infty {1 \over {2n+1}}(H_{2n}-\ln 2)z^{2n+1}-z \ln 2,  \eqno(168a)$$
and
$$\mbox{Li}_2\left({{1 \pm z} \over 2}\right)={{(1\mp z)} \over 2}\sum_{\ell=0}^\infty
\sum_{n=\ell}^\infty {{H_n^{(2)}} \over 2^n}{n \choose \ell} (\pm z)^\ell.  \eqno(168b)$$
Equation (168a) follows from known series expansions of $\ln$ or $\tanh^{-1}$
and $\ln^2$ \cite{grad} (pp. 44, 45, 51).  For Eq. (168b), we use the generating
function relation (44) at $r=2$, perform binomial expansion, and reorder the
resulting double sum.

More recent treatments of three-electron integrals \cite{pach} give analytic 
results that contain differences of dilogarithms, and these may often be written 
in terms of Clausen Cl$_2$ values.

In their derivation of the three-electron generating integral $I$, Fromm and Hill
\cite{fromm} (pp. 1015-16) pointed out that it possesses the larger symmetry of the
symmetric group $S_4$, that is, the permutation group on four objects. This
symmetry was in fact key in the discovery of their final general formula for $I$.

\medskip
\centerline{\bf Concluding brief discussion}
\medskip

For real $h \notin (-1,0)$, it would be desirable to have a general expression 
for integrals of the form
$$J(c,d,h)=\int_0^1 {1 \over {(x+h)}}{{\ln(x+c)} \over \sqrt{d^2-x^2}}dx, 
~~~~~d >1,  ~~~~ c/d > 1, \eqno(169)$$
solely in terms of Clausen function Cl$_2$ and elementary function values.
It should be emphasized, however, that much less than the general case of
Eq. (161) is required in order to alternatively evaluate Eq. (156).  This is
because not only strictly $d=\sqrt{3}$, but all three
parameters $c$, $d$, and $h$ are very tightly related.  There is no doubt
that integrals such as $J(3,\sqrt{3},\pm 2)$ may be evaluated in terms of
dilogarithms and logarithms--this can be shown in a number of ways.  
The remaining tasks in this approach would be to demonstrate that these dilogarithms reduce to Cl$_2$ values and that the overall result for Eq. (157) agrees with Eq. (3).

We have given a number of example logarithmic-trigonometric integrals that
evaluate in terms of Cl$_2$ and logarithms (or, their equivalents as inverse
trigonometric functions), including those of Appendix D.  Still others can
be found in the literature, by suitable change of variable.  Just one brief
example is provided by the integrals of Ref. \cite{lunev}.

In our treatment, we have very purposely avoided introducing two-variable
hypergeometric functions, in particular the Appell $F_1$ and $F_3$ functions.
Many of the integrals that we have considered are indeed expressible in terms
of these functions and their parametric derivatives.  While this may well be
an area worthy of further investigation, the advantages as regards an
efficient and transparent evaluation of Eq. (156) are not readily apparent.

In the course of our investigation we developed many other integral and 
summation representations.  Therefore, several of the stated results should
be taken as representative, and not exhaustive.  
The results such as we have presented strengthen the relations between 
Feynman diagram integrals, special functions, analytic number theory, and
hyperbolic geometry.  We have proved both of the two major conjectures of
Ref. \cite{broadhurst}.  In so doing, we have noted a sort of complementarity
between Feynman diagrams at the three-loop level, and the important three-electron
integrals of atomic physics.

\pagebreak
\centerline{\bf Acknowledgements}
This work was partially supported by Air Force contract number FA8750-06-1-0001.
I thank N. Lubbers for useful discussion.

\pagebreak
\centerline{\bf Appendix A:  Alternative evaluation of the sum $S(2,\beta)$}

Herein we re-express the sum $S(2,\beta)$ of Eq. (4) and find again the special
case of Eq. (5).  

We make use of an extension of the following result of Ramanujan (\cite{berndt}, 
Entry 12, p. 257).  Let
$$H(x) = \sum_{k=1}^\infty {H_k \over {2k-1}}x^{2k-1}, ~~~~~~~~|x| < 1, \eqno(A.1)$$
where $H_k$ is the $k$th harmonic number.  Then for $0 < x < 1$, $H$ satisfies
$$H\left({{1-x} \over {1+x}}\right)=(\ln 2-1)\ln x+{{1+x} \over {1-x}}\ln\left(
{{4x} \over {(1+x)^2}}\right)+{1 \over 4}\ln^2 x+{\pi^2 \over {12}}+\mbox{Li}_2
(-x).  \eqno(A.2)$$
We extend the range of validity of Eq. (A.2).  Given this result for $x\in (0,1)$,
since the logarithm and dilogarithm functions may be analytically continued to
the whole complex plane, so too may the function $H(x)$.

Since $H_{k+1}=H_k+1/(k+1)$, we first have from Eq. (A.1)
$$H(x)-2\tanh^{-1}x-{1 \over x}\ln(1-x^2)=\sum_{k=0}^\infty {H_k \over {2k+1}}
x^{2k+1}.  \eqno(A.3)$$
From Eq. (B.2) we obtain
$$H(y)=(\ln 2-1)\ln\left({{1-y} \over {1+y}}\right)+{1 \over y}\ln(1-y^2)
+{1 \over 4}\ln^2\left({{1-y} \over {1+y}}\right)+{\pi^2 \over {12}}+\mbox{Li}_2
\left({{y-1} \over {y+1}}\right).  \eqno(A.4)$$
From these two equations we therefore have
$$\sum_{k=0}^\infty {H_k \over {k+1/2}} x^{2k}={2 \over x}\left[
\ln 2\ln\left({{1-x} \over {1+x}}\right)+{1 \over 4}\ln^2\left({{1-x} \over {1+x}}\right)+{\pi^2 \over {12}}+\mbox{Li}_2\left({{x-1} \over {x+1}}\right)
\right].  \eqno(A.5)$$
If we choose $x=\pm i/\beta^{3/2}$, this sum is $S(2,\beta)$.  

We now specialize to $x=i/2\sqrt{2}$, and use the relation (114).  We have the
relations
$$\mbox{Li}_2\left({{x-1} \over {x+1}}\right)=\mbox{Li}_2[e^{i(\theta_++\pi)}]
={\pi^2 \over 6}-{1 \over 4}(\pi^2-\theta_+^2)+i\mbox{Cl}_2(\theta_++\pi),$$ 
$$\ln\left({{1-x} \over {1+x}}\right)=-2i\tan^{-1}\left({1 \over {2 \sqrt{2}}}
\right), \eqno(A.6)$$
where $\theta_+$ is defined just above Eq. (4) of the text and we recall that
$\omega=\tan^{-1}(1/2\sqrt{2})=\sin^{-1}(1/3)$.  Furthermore, by elementary
trigonometry we have $\theta_+=-2\omega$.
Therefore, we recover the specific relation (5).

{\bf Remark}.  Any time that $x$ in Eq. (A.5) is pure imaginary, meaning that the
sum $S(2,\beta)$ has sign alternation in the summand, this sum may be written
in terms of a sole Clausen function value.  
See Appendix C for the explicit result for general real values of $\beta$.

\pagebreak
\centerline{\bf Appendix B:  Relation of $\Phi(z,2,a)$ to the $_3F_2$ function}

An integral representation of the Lerch zeta function is contained in
$$\Phi(z,s,a)=\sum_{n=0}^\infty {1 \over {(n+a)^s}}={{(-1)^{s-1}} \over {\Gamma(s)}}
\int_0^1 {u^{a-1} \over {1-zu}}\ln^{s-1} u ~du, ~~~~~~\mbox{Re} ~s>1.   \eqno(B.1)$$
Therefore in particular we have
$$\Phi\left(-{1 \over 8},2,{1 \over 2}\right)=-\int_0^1 {1 \over \sqrt{u}}{{\ln u}
\over {(1+u/8)}}du, \eqno(B.2)$$
with the integral being an alternative to Eq. (29).  

For positive integers $k$ it is easy to see that
$$\Phi(z,k,a)=a^{-k} ~_{k+1}F_k(1,a,\ldots,a;a+1,\ldots,a+1;z), \eqno(B.3)$$
where $_pF_q$ is the generalized hypergeometric function.
Therefore, we have for the value considered in Lemma 2,
$$\Phi\left(-{1 \over 8},2,{1 \over 2}\right)=4 ~_3F_2\left(1,{1 \over 2},
{1 \over 2};{3 \over 2},{3 \over 2};-{1 \over 8}\right).  \eqno(B.4)$$
By using a result of Rainville \cite{rainville} (Theorem 38) we may derive
the value (B.4) by integrating over a certain $_2F_1$ function.  For Re $\alpha>0$
and Re $\beta>0$ we have
$$_3F_2(a_1,a_2,\alpha;b_1,\alpha+\beta;ct)={t^{1-\alpha-\beta} \over {B(\alpha,\beta)}}
\int_0^t x^{\alpha-1}(t-x)^{\beta-1} ~_2F_1(a_1,a_2;b_1;cx) ~dx,  \eqno(B.5)$$
where $B(x,y)=\Gamma(x)\Gamma(y)/\Gamma(x+y)$ is the Beta function.  Therefore,
with $\alpha=1/2$ and $\beta=1$, we have
$$_3F_2\left(1,{1 \over 2},{1 \over 2};{3 \over 2},{3 \over 2},ct\right)
={t^{1/2} \over 2}\int_0^t x^{-1/2} ~_2F_1\left({1 \over 2},1;{3 \over 2};cx
\right)dx$$
$$={t^{1/2} \over {2\sqrt{c}}}\int_0^t {1 \over x} \tanh^{-1}(\sqrt{cx}) ~dx,
\eqno(B.6)$$
and we see the reappearance of the integral of Eq. (32).  We obtain
$$_3F_2\left(1,{1 \over 2},{1 \over 2};{3 \over 2},{3 \over 2},ct\right)
={1 \over {4\sqrt{ct}}}\left[4 \mbox{Li}_2(\sqrt{ct})-\mbox{Li}_2(ct)\right],
\eqno(B.7)$$
wherein the relation $\chi_2(z)=\mbox{Li}_2(z)-{1 \over 4}\mbox{Li}_2(z^2)$
holds.  We may also recall that Ti$_2(y)=-i \chi_2(iy)$ for $y \in R$, where
Ti is the arctangent integral
$$\mbox{Ti}_2(x)=\int_0^x {{\tan^{-1} t} \over t} dt.  \eqno(B.8)$$

By using Eqs. (28), (31), and (B.1) we may summarize relations between a set of
special functions:
$$_3F_2\left({1\over 2},{1 \over 2},1;{3 \over 2},{3 \over 2};z\right)
=\sum_{j=0}^\infty {{(1/2)_j^2} \over {(3/2)_j^2}} z^j=\sum_{j=0}^\infty {z^j
\over {(2j+1)^2}}$$
$$={1 \over \sqrt{z}} \chi_2(\sqrt{z})={1 \over 4}\Phi\left(z,2,{1 \over 2}\right)
={1 \over {2\sqrt{z}}}\left[\mbox{Li}_2(\sqrt{z})-\mbox{Li}_2(-\sqrt{z})\right]$$
$$={1 \over 2}\left[_3F_2(1,1,1;2,2;\sqrt{z})+ ~_3F_2(1,1,1;2,2;-\sqrt{z})\right]$$
$$=-{1 \over 4}\int_0^1 {1 \over \sqrt{u}} {{\ln u ~du} \over {(1-zu)}}.  \eqno(B.9)$$

We may also mention
$$\int_0^1 \ln x\ln(1-zx^2)dx=2-{2 \over \sqrt{z}}\tanh^{-1} \sqrt{z}-{z \over 2}
\Phi(z,2,3/2)-\ln(1-z), \eqno(B.10)$$
being a $z \neq 1$ generalization of \cite{grad} (p. 559),
where $\Phi(z,2,3/2)={1 \over z}[\Phi(z,2,1/2)-4]$.

\pagebreak
\centerline{\bf Appendix C:  General expression for $S(2,\beta)$}

Here we evaluate the sums of Eq. (4) for $\alpha=2$.  We have
{\newline \bf Proposition C1}.  Let $\beta \in R$ and put
$$\theta=\cos^{-1}\left({{1-\beta^3} \over {1+\beta^3}}\right).  \eqno(C.1)$$
Then we have
$$S(2,\beta)=2\beta^{3/2}[\mbox{Cl}_2(\theta) -2 \cot^{-1} \beta^{3/2} \ln 2].
\eqno(C.2)$$
While this result follows directly from Lemma 2(b) of Ref. \cite{coffey07},
we prove it in another way.  In so doing, we find interesting intermediate
relations.  Indeed, we shall then have
{\newline \bf Proposition C2}.  For $\beta \in R$, and $\theta$ as in Eq. (C.1),
$$\beta^{3/2}\mbox{Cl}_2(\theta)=\sum_{k=1}^\infty {1 \over k}{1 \over {(2k-1)}}
~_2F_1\left(1,k;k+1;-{1 \over \beta^3}\right).  \eqno(C.3)$$

We begin by inserting the partial fractions form of the digamma function 
(e.g., \cite{nbs}, p. 259) into Eq. (4).  This gives
$$S(\alpha,\beta)=\sum_{n=1}^\infty {{(-1/\beta^3)^n} \over {n+1/\alpha}}
n\sum_{k=1}^\infty {1 \over {k(k+n)}}$$
$$=\sum_{k=1}^\infty {1 \over k}\sum_{n=1}^\infty \left(-{1 \over \beta^3}\right)^n 
{n \over {(n+1/\alpha)}} {1 \over {(k+n)}}, \eqno(C.4)$$
where we interchanged sums.  Performing the inner sum then gives
$$S(\alpha,\beta)={\alpha \over \beta^3}\sum_{k=1}^\infty {1 \over k}
{1 \over {(\alpha k-1)}}\left[{1 \over {(1+\alpha)}} ~_2F_1\left(1,1+{1 \over \alpha};2+{1 \over \alpha};-{1 \over \beta^3}\right)\right.$$
$$\left.-{k \over {(k+1)}} ~_2F_1\left(1,k+1;k+2;-{1 \over \beta^3}\right)\right].
\eqno(C.5)$$

We now specialize to $\alpha=2$, finding
$$S(2,\beta)=-4 \beta^{3/2} \ln 2 \cot^{-1} \beta^{3/2}+2\sum_{k=1}^\infty {1 \over k}
{1 \over {(2k-1)}} ~_2F_1\left(1,k;k+1;-{1 \over \beta^3}\right).  \eqno(C.6)$$
We now focus on the $_2F_1$ function in this equation, that we transform in the following ways.  By \cite{grad} (p. 1043) and then the use of the integral representation of the hypergeometric function \cite{grad} (p. 1040) we obtain
$$_2F_1\left(1,k;k+1;-{1 \over \beta^3}\right)=\left(1+{1 \over \beta^3}\right)^{-1}
~_2F_1\left(1,1;k+1; {1 \over {1+\beta^3}}\right)$$
$$=\left(1+{1 \over \beta^3}\right)^{-1}k \int_0^1 (1-t)^k \left(1-{t \over 
{1+\beta^3}}\right)^{-1} dt, \eqno(C.7)$$
wherein we used $1/B(1,k)=k$ where $B$ is the Beta function.  
Substituting Eq. (C.7) into the sum on the right side of Eq. (C.6) we have
$$\sum_{k=1}^\infty {1 \over k}{1 \over {(2k-1)}} ~_2F_1\left(1,k;k+1;-{1 \over \beta^3}\right)=\left(1+{1 \over \beta^3}\right)^{-1} \sum_{k=1}^\infty {1 \over {(2k-1)}} \int_0^1 (1-t)^k \left(1-{t \over {1+\beta^3}}\right)^{-1} dt$$
$$=\left(1+{1 \over \beta^3}\right)^{-1} \int_0^1 \sqrt{1-t} \tanh^{-1} \sqrt{1-t}
\left(1-{t \over {1+\beta^3}}\right)^{-1} dt, \eqno(C.8)$$
using a well known series for arc tanh (e.g., \cite{grad}, p. 51).

By the change of variable $t=1-v^2$, we find
$$\sum_{k=1}^\infty {1 \over k}{1 \over {(2k-1)}} ~_2F_1\left(1,k;k+1;-{1 \over \beta^3}\right)=2\beta^3 \int_0^1 {{v^2 \tanh^{-1} v} \over {v^2+\beta^3}}dv$$
$$=2 \beta^3\left[\ln 2-\beta^3 \int_0^1 {{\tanh^{-1} v} \over {v^2+\beta^3}}dv
\right].  \eqno(C.9)$$
This equation may be evaluated in terms of dilogarithms of complex argument by means 
of Eqs. (3.12.7) and (3.12.11) of Ref. \cite{duke}, by writing the denominator
of the integrand as $(v-v_+)(v-v_-)$ where $v_\pm = \pm i \beta^{3/2}$.  However,
it is more expedient for our current purpose to continue with the changes of
variable $y=v/\beta^{3/2}$ and $u=1/y$, leading to 
$$\int_0^1 {{\tanh^{-1} v} \over {v^2+\beta^3}} dv
={1 \over {2 \beta^{3/2}}} \int_{\beta^{3/2}}^\infty \ln\left({{u+\beta^{3/2}} \over
{u-\beta^{3/2}}}\right) {{du} \over {u^2+1}}$$
$$={1 \over {2 \beta^{3/2}}}\mbox{Cl}_2(\theta).  \eqno(C.10)$$  
In the last step we applied the integral representation of Lemma 2(a) of Ref. 
\cite{coffey07}.  By collecting terms, we then have Proposition C1 and by Eq. (C.6)
we deduce Proposition C2.

We indicate another method by which to obtain the general expression (C.2).
By formula (4.2.4) of Ref. \cite{duke} (p. 30), we have the integral representation
for harmonic numbers
$$H_n=-n \int_0^1 x^{n-1} \ln (1-x) ~dx.  \eqno(C.11)$$
Using this representation, interchanging summation and integration, we have
$$\sum_{n=0}^\infty {{(-x)^n} \over {(n+1/2)}}H_n=-\int_0^1\ln(1-t)
\left[{1 \over {t(1+xt)}}-{{\tan^{-1} \sqrt{xt}} \over {\sqrt{x} t^{3/2}}} ~\right]dt.
\eqno(C.12)$$
The integral may be evaluated in terms of the dilogarithmic difference
$${i \over \sqrt{x}}\left[\mbox{Li}_2\left({{i+\sqrt{x}} \over {-i+\sqrt{x}}}\right)-
\mbox{Li}_2\left({{-i+\sqrt{x}} \over {i+\sqrt{x}}}\right)\right].$$
This quantity is easily written in terms of Clausen function values using the
angle $2\theta=2\tan^{-1} \sqrt{x}$.  Once again, at $x=1/\beta^3=1/8$, we have
the appearance of the angle $\tan^{-1}(1/2\sqrt{2})=\omega=-\theta_+/2$ in $S(2,2)$.

\pagebreak
\centerline{\bf Appendix D:  A two-parameter Clausen function integral}

For $t$ and $y$ in $R$, let
$$\omega_1 \equiv \tan^{-1}\left({{e^y \sin t} \over {1+e^y \cos t}}\right), ~~~~~~
\omega_3 \equiv \tan^{-1}\left({{e^y \sin t} \over {1-e^y \cos t}}\right).
\eqno(D.1)$$
We have
{\newline \bf Proposition D1}.  
$$\int_0^y {{x \cosh x ~dx} \over {\cosh 2x-\cos 2t}}={{\csc t} \over 4}
\left[-4\mbox{Cl}_2(\pi+t)-\mbox{Cl}_2(2\omega_1)+\mbox{Cl}_2(2(\omega_1-t))\right.$$
$$\left.+\mbox{Cl}_2 (2t)-\mbox{Cl}_2(2\omega_3)+\mbox{Cl}_2(2(\omega_3+t))\right].  \eqno(D.2)$$

Define the integrals
$$I_\pm(y,t)=\int_0^y {{x ~dx} \over {\cosh x\pm \cos t}}.  \eqno(D.3)$$
As part of our proof of Proposition D1 we find the separate evaluations
$$I_+(y,t)=\csc t[2\mbox{Cl}_2(t)-\mbox{Cl}_2(2\omega_1)+\mbox{Cl}_2(2(\omega_1-t))],
\eqno(D.4)$$
and
$$I_-(y,t)=\csc t[2\mbox{Cl}_2(\pi-t)-\mbox{Cl}_2(2\omega_3)+\mbox{Cl}_2 (2(\omega_3+t))]. \eqno(D.5)$$
Proposition D1 gives a direct evaluation of formula 3.532.2 of Ref. \cite{grad}
(p. 355) in terms of Clausen function values.  It gives, as we briefly describe
below, Eq. (132) as a limiting special case.  Equation (D.5) provides the
equivalent of formula 3.531.8 of Ref. \cite{grad} (p. 354), and the limit as $y \to
\infty$ of Eq. (D.4) gives the equivalent of formula 3.531.2 there.

Proof.  Since $\cosh 2x-\cos 2t=2\sinh^2x+1-(2\cos^t-1)=2(\sinh^x -\cos^2 t+1)
=2(\cosh^2 x-\cos^2 t)$, we have
$$\int_0^y {{x \cosh x ~dx} \over {\cosh 2x-\cos 2t}}={1 \over 2}\int_0^y
{{x \cosh x ~dx} \over {(\cosh x-\cos t)(\cosh x+\cos t)}}$$
$$={1 \over 4}\int_0^y x\left[{1 \over {\cosh x- \cos t}}+{1 \over {\cosh x+ \cos t}}\right]dx={1 \over 4}[I_-(y,t)+I_+(y,t)].  \eqno(D.6)$$
We may write, with $u=\cosh x$,
$$I_\pm(y,t)=\int_1^{\cosh y} {{\ln(u+\sqrt{u^2-1})} \over {(u \pm \cos t)
\sqrt{u^2-1}}} du.  \eqno(D.7)$$
We now put $z=u+\sqrt{u^2-1}$, yielding
$$I_\pm(y,t)=2 \int_1^{e^y} {{\ln z ~dz} \over {z^2\pm 2z \cos t+1}}=2\int_1^{e^y}
{{\ln z ~dz} \over {(z-z_+^{(\pm)})(z-z_-^{(\pm)})}}.  \eqno(D.8)$$
These integrals are next evaluated in terms of dilogarithms:
$$I_\pm(y,t)={{-2} \over {(z_+-z_-)}}\left\{y\left[\ln\left(1-{e^y \over z_-}\right)
-\ln\left(1-{e^y \over z_+}\right)\right]+\mbox{Li}_2\left({1 \over z_+}\right)-
\mbox{Li}_2\left({1 \over z_-}\right)\right.$$
$$\left. +\mbox{Li}_2\left({e^y \over z_-}\right)-\mbox{Li}_2\left({e^y \over z_+}\right)\right\}, \eqno(D.9)$$
where, for $I_+$, $z_\pm=-e^{-it}, -e^{it}$, and for $I_-$, $z_\pm=e^{it}, e^{-it}$,
so that $z_+-z_-=2i \sin t$ in both cases.  The arguments of the dilogarithms of this equation are written in polar form and  Eqs. (29) and (30) are applied.  Omitting details, we determine, for instance, for $I_+$,
$$\mbox{Im}\left[\mbox{Li}_2\left({e^y \over z_-}\right)-\mbox{Li}_2\left({e^y \over z_+}\right)\right]=2\omega_1 y+\mbox{Cl}_2(2\omega_1)-\mbox{Cl}_2(2(\omega_1-t))-
\mbox{Cl}_2(2t), \eqno(D.10)$$
and for $I_-$,
$$\mbox{Im}\left[\mbox{Li}_2\left({e^y \over z_-}\right)-\mbox{Li}_2\left({e^y \over z_+}\right)\right]=2\omega_3 y+\mbox{Cl}_2(2\omega_3)-\mbox{Cl}_2(2(\omega_1+t))+
\mbox{Cl}_2(2t). \eqno(D.11)$$
In writing such equations we have used the $2\pi$ periodicity and oddness of the
function Cl$_2$.  We further find that the logarithm terms of Eq. (D.9) are
cancelled by the $2\omega_1 y$ and $2\omega_3 y$ terms of Eqs. (D.10) and (D.11).
The additional use of the duplication formula (37) leads to Eqs. (D.4) and (D.5)
and hence, by Eq. (D.6), Eq. (D.2), completing the Proposition.

{\bf Remarks}.  Alternatively in the proof, we could have applied the integral
representation (140) for the evaluation of Eq. (D.8).

We have as $y \to \infty$ in Eq. (D.1), $\omega_1 \to t$ and $\omega_3 \to -t$, 
giving
$$I_+(y,t) \to 2\csc t ~\mbox{Cl}_2(\pi-t), ~~~~~
I_-(y,t) \to 2\csc t ~\mbox{Cl}_2(t), ~~~~~~~~ y \to \infty.  \eqno(D.12)$$
Therefore, we have
$${1 \over 4}(I_++I_-) \stackrel{y\to \infty}{\to} {{\csc t} \over 2}
[\mbox{Cl}_2(t)+\mbox{Cl}_2(\pi-t)]={{\csc t} \over 2}\left[{1 \over 2}\mbox{Cl}_2(2t)+2\mbox{Cl}_2(\pi-t)\right]$$
$$={{\csc t} \over 2}\left[2 \mbox{Cl}_2(t)-{1 \over 2}\mbox{Cl}_2(2t)\right],
\eqno(D.13)$$
and Eq. (132), the equivalent of formula 3.533.1 \cite{grad} (p. 355), is
recovered.

Upon the change of variable $v=1/u$, we have
$$I_\pm(y,t)=\int_{\mbox{\tiny{sech}} y}^1 {{[-\ln v+\ln(1+\sqrt{1-v^2})]} \over {(1 \pm \cos t ~v) \sqrt{1-v^2}}} dv.  \eqno(D.14)$$
So, with $v=\cos \theta$, we have obtained other log trigonometric integrals:
$$I_\pm(y,t)=\int_0^{\cos^{-1} \mbox{\tiny{sech}} y} {{[\ln(1+\sin \theta)-\ln \cos \theta]} \over {1 \pm \cos t \cos \theta}}d \theta.  \eqno(D.15)$$

\pagebreak


\begin{thebibliography}{99}
\bibitem{nbs}M. Abramowitz and I. A. Stegun,
{Handbook of Mathematical Functions, National Bureau of Standards (1972).}
\bibitem{arnold}P. Arnold and L. G. Yaffe,
{$\epsilon$ expansion analysis of very weak first-order transitions in the cubic
anisotropy model.  I, Phys. Rev. D {\bf 55}, 7760-7775 (1997).}
\bibitem{berndt}B. C. Berndt,
{Ramanujan's notebooks, Part I, Springer (1985).}
\bibitem{broadhurst}D. J. Broadhurst,
{A dilogarithmic 3-dimensional Ising tetrahedron, arxiv/hep-th/9805025 v3 (1998);
Eur. Phys. J. C {\bf 8}, 363-366 (1999).}
\bibitem{broad2}D. J. Broadhurst,
{Solving differential equations for 3-loop diagrams: relation to hyperbolic
geometry and knot theory, arxiv/hep-th/9806174v2 (1998).} 
\bibitem{choy}T. C. Choy, D. Sherrington, M. Thomsen, and M. F. Thorpe,
{Local magnetic field distributions.  II.  Further results, Phys. Rev. B {\bf 31}, 7355-7367 (1985).}
\bibitem{coffey07}M. W. Coffey,
{Evaluation of a $\ln \tan$ integral arising in quantum field theory, preprint 
arXiv/math-ph/0801.0272v1 (2008).}
\bibitem{coxeter}H. S. M. Coxeter,
{The functions of Schl\"{a}fli and Labatschefsky, Quart. J. Math. (Oxford) {\bf 6}
13-29 (1935).}
\bibitem{ded}P. J. de Doelder,
{On the Clausen integral Cl$_2(\theta)$ and a related integral, J. Comput. Appl. Math. {\bf 11}, 325-330 (1984).}
\bibitem{duke}A. Devoto and D. W. Duke,
{Table of integrals and formulae for Feynman diagram calculations. Riv. Nuovo Cim.
{\bf  7}, 1-39 (1984).}
\bibitem{espinosa}J. R. Espinosa and R.-J. Zhang,
{Complete two-loop dominant corrections to the mass of the lightest {\cal CP}-even Higgs boson in the minimal supersymmetric standard model, Nucl. Phys. B {\bf 586},
3-38 (2000).}
\bibitem{fromm}D. M. Fromm and R. N. Hill,
{Phys. Rev. A {\bf 36}, 1013-1044 (1987).}
\bibitem{grad}I. S. Gradshteyn and I. M. Ryzhik,
{Table of Integrals, Series, and Products, Academic Press, New York (1980).}
\bibitem{grosjean}C. C. Grosjean,
{Formulae concerning the computation of the Clausen integral Cl$_2(\theta)$, 
J. Comput. Appl. Math. {\bf 11}, 331-342 (1984).}
\bibitem{hansen}E. R. Hansen,
{A table of series and products, Prentice Hall (1975).}
\bibitem{harris}F. E. Harris,
{Analytic evaluation of three-electron atomic integrals with Slater wave functions,
Phys. Rev. A {\bf 55}, 1820-1831 (1997).}
\bibitem{hold}J. T. Holdeman, Jr.,
{A method for the approximation of functions defined by formal series expansions in
orthogonal polynomials, Math. Comp. {\bf 23}, 275-287 (1969).  Note that in Eq. (52)
of this reference, $(2n-1)$ should read $(2n+1)$, as given in \cite{hansen},
formula (46.2.21).}
\bibitem{lewin}L. Lewin,
{Polylogarithms and associated functions, North Holland (1981).}
\bibitem{herbert}E. Herbert Li and S. K. Tin,
{The Lobachevskiy's function and related integrals, Inst. Math. Appls. Bull. {\bf 27},
175-180 (1991).}
\bibitem{lunev}F. A. Lunev,
{Evaluation of two-loop self-energy diagram with three propagators,
Phys. Rev. D {\bf 50}, 7735-7737 (1994).  It appears that in Eq. (18), the
integrand factor $(1-\lambda y)$ should read $(1-\lambda x)$.}
\bibitem{milnor}J. Milnor,
{Hyperbolic geometry:  the first 150 years, Bull. Amer. Math. Soc. {\bf 6},
9-24 (1981).}
\bibitem{nielsen}N. Nielsen, 
{Nova Acta (Leopold), {\bf 90}, 154-155 (1909).}
\bibitem{pach}K. Pachucki and M. Puchalski,
{Extended Hylleraas three-electron integral, Phys. Rev. A {\bf 71}, 032514 (2005);
K. Pachucki, M. Puchalski, and E. Remiddi, Recursion relations for the generic
Hylleraas three-electron integral, Phys. Rev. A {\bf 70}, 032502 (2004); 
K. Pachucki and J. Komasa, Three-electron integral in a Gaussian basis set with 
linear terms, Phys. Rev. A {\bf 70}, 022513 (2004).}
\bibitem{rainville}E. D. Rainville,
{Special functions, Macmillan (1960).}
\bibitem{raj}A. K. Rajantie,
{Feynman diagrams to three loops in three-dimensional field theory, Nucl. Phys. B
{\bf 480}, 729-752 (1996).}
\bibitem{remiddi}E. Remiddi,
{Analytic value of the atomic three-electron correlation integal with Slater wave
functions, Phys. Rev. A {\bf 44}, 5492-5501 (1991); F. E. Harris et al., Phys. Rev. 
A {\bf 69}, 056501 (2004); J. S. Sims and S. A. Hagstrom, Phys. Rev. A {\bf 68},
016501 (2003).}
\bibitem{sri}H. M. Srivastava and J. Choi,
{Series associated with the zeta and related functions, Kluwer Academic (2001).}
\bibitem{wannier}G. H. Wannier,
{Antiferromagnetism.  The triangular Ising net, Phys. Rev.  {\bf 79}, 357-364 (1950), Phys. Rev. B {\bf 7} Errata, 5017 (1973).}
\end{thebibliography}
\end{document}